 \documentclass[final,3p,times,twocolumn]{elsarticle}

\usepackage{comment}
\usepackage{amssymb}
\usepackage{amsmath}
\usepackage{chemfig}
\usepackage{spalign}
\usepackage[version=4]{mhchem}
\usepackage{cleveref}
\usepackage{upgreek}
\usepackage{subfig}

\usepackage{dblfloatfix}
\journal{Powder Technology}

\begin{document}

\begin{frontmatter}



\title{Influence of the particle distribution on dust explosions in the 20~L sphere}


\author{Kasun Weerasekara$^\text{a,b,*}$} 
\author{Stefan H. Spitzer$^\text{c}$}
\author{Sabine Zakel$^\text{a}$}
\author{Holger Grosshans$^{\text{a,b}}$}
\affiliation{organization={Physikalisch-Technische Bundesanstalt},
            addressline={Bundesallee 100}, 
            city={Braunschweig},
            postcode={38116}, 
            country={Germany}}
\affiliation{organization={Otto von Guericke University of Magdeburg},
            addressline={Universitätspltz 2}, 
            city={Magdeburg},
            postcode={39106}, 
            country={Germany}}

\affiliation{organization={EIfI-Tech},
            addressline={Universitätspark 1/1}, 
            city={Schwäbisch Gmünd},
            postcode={73525}, 
            country={Germany}}

\begin{abstract}
It is essential to standardize the safety characteristics of dust explosions to mitigate their impact on the process industries. The 20 L sphere primarily investigates the safety characteristics, namely explosion pressure ($P_\text{ex}$) and the rate of pressure rise ($(dP/dt)_\text{ex}$), of dust explosions at the laboratory level. Ensuring uniform dust distribution inside the sphere is essential for accurate data acquisition and standardization. However, whirls created by the incoming flow through the nozzle yield particles to concentrate near the wall before ignition.
This study simulated the explosion inside a 20~L sphere to investigate the impact of near-wall particle concentration on the safety characteristics. The OpenFOAM model based on the Euler-Lagrangian approach was benchmarked against experimental data of lycopodium dust explosions. A novel radial homogeneity parameter $\Phi$ $(0\leq \Phi \leq 1)$ quantifies the near-wall particle concentration. The parameter $\Phi$ is calculated using a power law based on the radial component of particle coordinates, $\Phi = 1$ indicating a uniform distribution, and $\Phi = 0$ for all particles concentrating on the wall. Different particle distributions ($\Phi = 0.1,0.2,...,1$) are initiated before ignition. 
As $\Phi$ decreases from 1, $P_\text{ex}$ and $(dP/dt)_\text{ex}$ first decrease, but beyond a certain point, both parameters increase. At $\Phi=0.1$, both $P_\text{ex}$ and $(dP/dt)_\text{ex}$ reach their highest values, which are 1.75$\%$ and 10.1$\%$ higher than the uniform distribution, respectively. The lowest values arise at $\Phi = 0.7$, with reductions of 0.25$\%$ and 5.6$\%$ compared to the uniform distribution. Thus, high near-wall concentrations enhance explosion intensity, while moderate concentrations result in lower intensity than the uniform distribution. 
\end{abstract}




\end{frontmatter}



\section{Introduction}

Dust explosions in process industries are caused by the ignition of fine combustible particles suspended in air, as commonly seen in grain silos, flour mills, and coal mines. Between the years 1785 and 2012, over 2870 dust explosions occurred worldwide, killing more than 4800 people \cite{explHis}. Therefore, understanding and regulating dust explosions is essential to protect human lives and physical resources.

A dust explosion occurs when an oxidant, combustible dust, dispersion of the dust, confinement, and an ignition source are all present. Eliminating one or more of these required elements can prevent the initiation of a dust explosion. To minimize the harm of these events, it is necessary to evaluate the safety characteristics of the explosion. In the laboratory, the 20~L sphere provides the explosion's pressure ($P_\text{ex}$), and pressure rise $(dP/dt)_\text{ex}$, and the MIKE 3 apparatus the minimum ignition energy. This study focuses on the 20 L sphere.

\begin{figure}[t]
    \centering
    \subfloat[]{\includegraphics[width=0.5\textwidth]{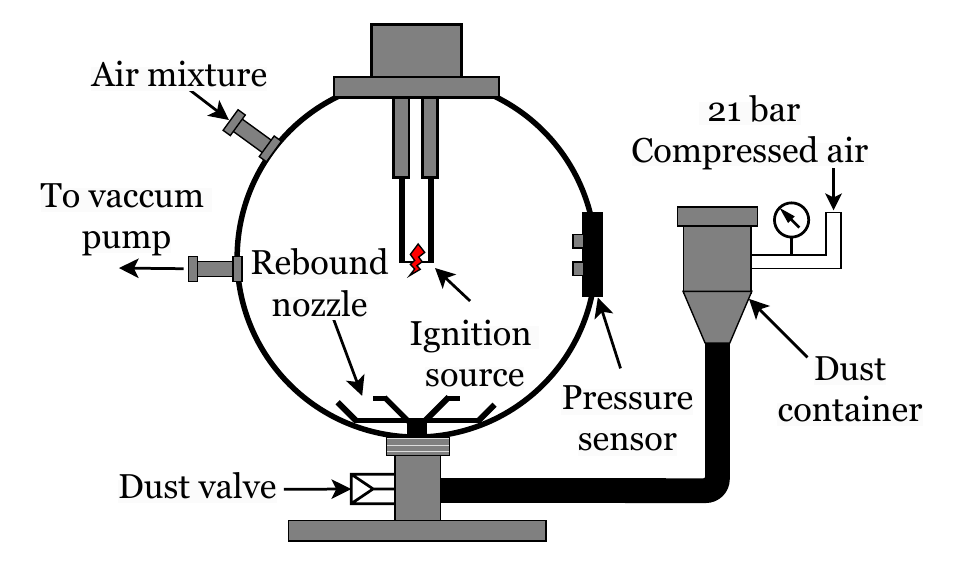}\label{fig:siwekschi}}\\
    \subfloat[]{\includegraphics[width=0.5\textwidth]{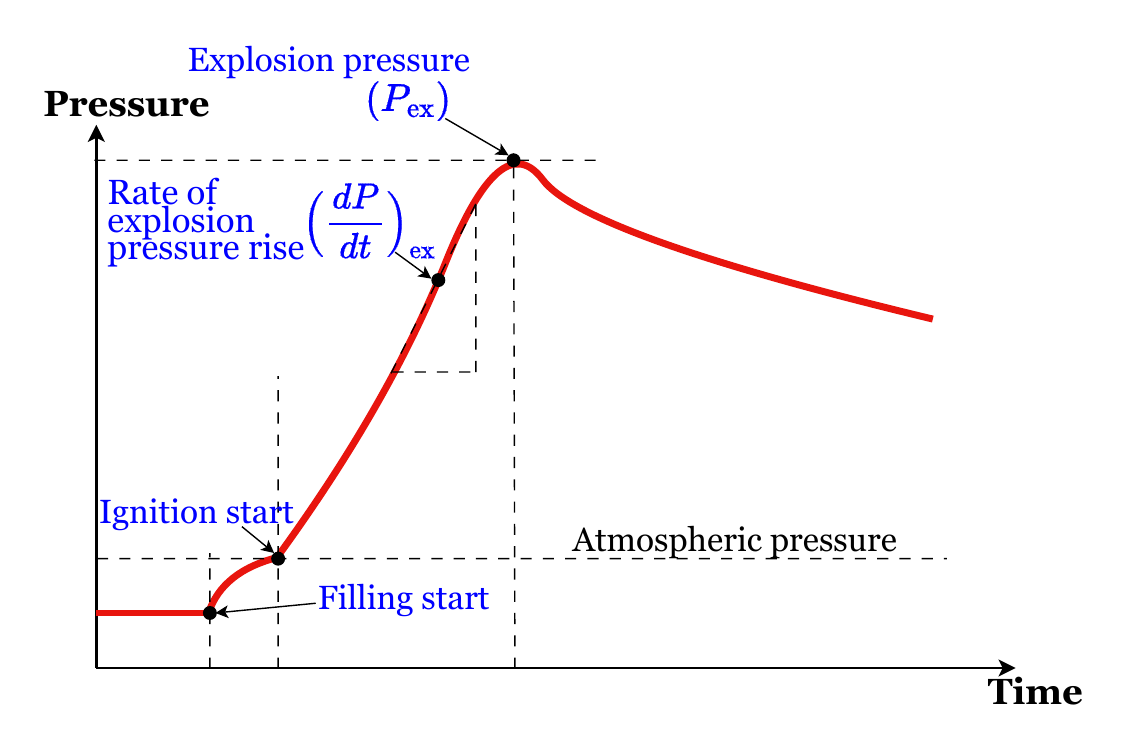}\label{fig:pressurePlotSiwek}}
    \caption{(a) Schematics of the 20~L sphere and (b) pressure evolution in the experiment.}
\end{figure}

The 20 L sphere contains two chemical igniters at its center, which ignite the explosive dust as it enters through a nozzle (Figure~\ref{fig:siwekschi}). Two internal pressure sensors measure the pressure versus time \mbox{(Figure~\ref{fig:pressurePlotSiwek})}. The initial pressure inside the sphere remains at 0.4 bar, below atmospheric pressure, and the dust container is pressurized up to 21 bar. Once the valve opens, dust particles flow through the nozzle into the sphere at high pressure. Ignition begins after dust particles fully disperse within the sphere. A brief waiting period, the ignition delay typically of 60 ms, ensures uniform dust dispersion. When ignition starts, the dust cloud explodes, and the pressure rapidly increases to a peak value before gradually decreasing. 

Accurately determining the dust's minimum explosible concentration (MEC) and other parameters requires a uniform particle distribution in the sphere \cite{cashdollar1993minimum}. To achieve a uniform distribution, selecting an appropriate nozzle is essential. However, the commonly used rebound nozzle does not produce a uniform dust distribution, and the particle concentration tends to be higher near the sphere walls.

\textcolor{black}{Cashdollar et al. \cite{cashdollar1993minimum}} measured local variations in dust concentrations using optical probes at different locations inside the sphere. A Siwek sphere with a setup of six optical probes, arranged in two sets of three probes and extending from the wall to the center along two perpendicular radial directions, revealed that dust particle concentrations were higher near the chamber walls than at the center \cite{KALEJAIYE201046}. Afterward, a transparent 20~L Siwek sphere enabled direct observation of the dust dispersion before ignition by shadowgraphy. Findings indicated a higher concentration of particles near the wall, especially around the rebound nozzle, suggesting that alternative nozzles could improve particle distribution homogeneity \cite{DU2015213}.

However, experimental setups have limitations in visualizing exact distributions inside the sphere. Well-validated 3D computational fluid dynamic (CFD) simulations of the 20L sphere can provide more details on the particles than experiments. For instance, a 3D CFD simulation using ANSYS-FLUENT showed that turbulent vortexes caused higher dust concentrations near the vessel walls \cite{DiBendettoCFD}. This occurs because particles first exit the nozzle and strike the wall, then move along the wall to the top of the sphere, falling and merging with the inflow from the nozzle. This movement induces two counter-rotating vortices inside the chamber. Eventually, the particles settle down, which reduces turbulence. 
\textcolor{black}{Di Sarli et al. \cite{DISARLI20148,Di_2013} presented through CFD simulations that particle distribution is influenced by both concentration and diameter, with high concentrations (500 g/m³) and larger diameters (250 $\upmu$m) causing particles to accumulate along vessel walls, while lower concentrations (100 g/m³) and smaller diameters (10 $\upmu$m) lead to particle accumulation at vortex boundaries and partial following of the fluid flow.} Further CFD analysis using OpenFOAM evaluated particle distribution for a new nozzle design and revealed that while the rebound nozzle resulted in high particle concentrations near the wall, the new design improved the concentration in the middle compared with the previous nozzle and a more even distribution throughout the sphere \cite{newdispOFWerner}.

In summary, experiments and simulations confirmed that particles in the sphere distributed inhomogeneously, with higher concentrations near the wall before the ignition. The present study investigated, through numerical simulations, how the particle \textcolor{black}{distribution} affects the safety characteristics. Section 2 details the computational model, and Section 3 presents the validation of the CFD model against experimental data. The analysis introduces a new parameter to evaluate the degree of homogeneity inside the chamber. Multiple simulations examine different levels of particle distribution, and Section 4 discusses the resulting explosion parameter changes and comparisons. Section 5 presents the study's conclusions and future directions.

\section{Computational Methodology}

This study applies numerical simulations along with experiments for benchmarking. This section discusses the computational methodology employed for the simulations. A dust explosion is a multiphysics and multiphase problem, including a fluid (air mixture) and a solid phase (dust particles). As a multiphase simulation technique, this study uses the Euler-Lagrangian approach, which couples the continuous phase in the Eulerian framework with the dispersed phase in the Lagrangian framework. This method is particularly effective for flows with a relatively low volume fraction of the dispersed phase, such as dust explosions.

To model the particle-fluid flow interactions and chemical reactions in dust explosions, we used the \texttt{coalChemistryFoam} solver of the OpenFOAM 2012 open-source package. This solver accounts for two-way coupling, i.e., the momentum transfer from the particles to the fluid and vice versa.

\subsection{Gas Phase Governing Equations}

The governing equations are based on Reynolds-Averaged Navier-Stokes (RANS) equations (\eqref{eqn:gas_mass} and \eqref{eqn:gas_momem}), along with the energy and species conservation equations (\eqref{eqn:gas_specie} and \eqref{eqn:gas_energy}),

\begin{equation}\label{eqn:gas_mass}
\frac{\partial \rho}{\partial t} + \nabla \cdot (\rho \mathbf{u}) = S_\text{M}\;,
\end{equation}

\begin{equation}\label{eqn:gas_momem}
\frac{\partial (\rho\mathbf{u})}{\partial t} + \nabla \cdot (\rho\mathbf{u} \otimes \mathbf{u}) = - \nabla {p} + \nabla \cdot \boldsymbol{\tau} - \nabla \cdot (\rho \mathbf{R}_f) + \rho \mathbf{g} + S_\text{F}\;, 
\end{equation}

\begin{equation}\label{eqn:gas_energy}
\begin{aligned}
\frac{\partial (\rho h)}{\partial t} + \frac{\partial (\rho k)}{\partial t}  + \nabla \cdot (\rho \mathbf{u} h)  +  \nabla \cdot (\rho \mathbf{u} k)  - \nabla \cdot (\alpha_\text{eff} \nabla h)  = \\ \frac{\partial p}{\partial t} 
 +\rho\mathbf{u} \cdot\mathbf{g}  + \dot{Q}_\text{react} 
 + \dot{Q}_\text{radi} + S_\text{E}\;,
\end{aligned}
\end{equation}

\begin{equation}\label{eqn:gas_specie}
\frac{\partial (\rho Y_k)}{\partial t} + \nabla \cdot (\rho \mathbf{u} Y_k) - \nabla \cdot \Bigl(D_\text{eff} \nabla(\rho Y_k)\Bigr) = \overline{\dot{\omega}}_k + S_{Y,k}\;.
\end{equation}

The perfect gas law, serving as the equation of state, 
\begin{equation}
\rho = \frac{p}{(R_\text{u}/W)T}
\end{equation}
closes the system of governing equations.

The source terms of the mass, momentum, energy, and species transport equations ($S_\text{M}$, $ S_\text{F}$, $S_\text{E}$, and $S_\text{Y k}$) couple the Lagrangian to the Eulerian equations. The source terms for the mass equation arise from moisture evaporation, devolatilization, and solid particle conversion. The source terms of the momentum equations link to the particles' drag. Ignition, radiative heating, and heat addition or absorption from chemical reactions form the source terms of the energy equation. In the species transport equation, reactants and products serve as source terms. The following chapters detail the modeling of these source terms.

\subsubsection{Homogeneous Chemical Reaction Modeling}

This section discusses the mathematical modeling of chemical reactions of the volatiles in the gas phase. 
This study adopts a four-step global mechanism to model the oxidation of \ce{CH4}, initially developed for hydrocarbon combustion \cite{JONES1988233}. This model implements 4 Arrhenius reactions, with two irreversible reactions and two reversible reactions:
\begin{center}
\begin{tabular}{@{}l@{\hskip 5pt}c@{\hskip 5pt}l@{\hskip 5pt}c@{\hskip 5pt}l@{\hskip 5pt}c@{\hskip 5pt}l@{}}
\ce{CH4} & + & \ce{0.5O2} & $\longrightarrow$ & \ce{CO} & + & \ce{2H2} \\
\ce{CH4} & + & \ce{H2O} & $\longrightarrow$ & \ce{CO} & + & \ce{3H2} \\
\ce{CO} & + & \ce{H2O} & $\longleftrightarrow$ & \ce{CO2} & + & \ce{H2} \\
\ce{H2} & + & \ce{0.5O2} & $\longrightarrow$ & \ce{H2O} & & \\
\end{tabular}
\end{center}
The kinetic rates for each reaction are provided by Islas et al.~\cite{ISLAS2022791}.

As an example, the reaction modeling is explained for a one-step methane combustion mechanism (CH$_4$+2O$_2$$\rightarrow$CO$_2$+2H$_2$O).
This mechanism is considered an `Irreversible Arrhenius Reaction,' a temperature-based reaction rate. For methane combustion,
\begin{equation}\label{eqn:kf}
    k_\text{f} = AT^{\beta}\text{exp}(-T_\text{a}/T)
\end{equation} 
calculates the reaction rate constant with Arrhenius reaction constants \mbox{$A=7\times10^{6}$mol\,m$^{-3}$s$^{-1}$},  $\beta=0$, \text{and} $T_\text{a}=\text{10 063} \text{ K}$  \cite{poinsot2005theoretical}.

Generally,
\begin{equation}\label{eqn:setChemical}
    \sum_{k=1}^{K} \nu_{k,i}'\mathcal{X}_{k}  \rightleftharpoons  \sum_{k=1}^{K} \nu_{k,i}''\mathcal{X}_{k} \;\;\;\; i = 1,2..I
\end{equation}
expresses the set of $I$ reactions with $K$ species. Here, $\nu_{k,i}'$, $\nu_{k,i}''$, and $\mathcal{X}_{k}$ are the forward and backward stoichiometric coefficients and the chemical symbol of the species $k$, respectively. They become for methane combustion, $I = 1$, $\nu_{k,i}' = \begin{bmatrix} 1 & 2 & 0 & 0\end{bmatrix}$, $\nu_{k,i}'' = \begin{bmatrix} 0 & 0 & 1 & 2\end{bmatrix}$, 
and $\mathcal{X}_{k} =\begin{bmatrix}\ce{CH4} & \ce{O2} & \ce{CO2} & \ce{H2O} \end{bmatrix}^{T}$.

The chemical reaction rate ($\dot{\omega}$) describes the velocity of a chemical reaction. It represents the change in the concentration of species for the $i^{th}$ reaction, expressed as,
\begin{equation}
    \dot{\omega}_{i} = k_{\text{f,}i}\prod_{k}^{} \bigr[\mathcal{X} \bigr]^{\nu_{k,i}'}_{k} - k_{\text{r},i}\prod_{k}^{} \bigr[\mathcal{X} \bigr]^{\nu_{k,i}''}_{k},
\end{equation}
where $k_{\text{f,}i}$, $k_{\text{r},i}$, and $\bigr[\mathcal{X}_{k} \bigr]$ are the forward and backward kinetic rates and the concentration of specie $k$, respectively. Methane combustion has no backward reaction ($k_{\text{r},i} = 0$). Thus,
\begin{equation}
    \dot{\omega}_\text{1} = k_{\text{f,}i=1}\bigr[\ce{CH4} \bigr]^{1} \bigr[\ce{O2} \bigr]^{2}\quad.
\end{equation}

The reaction rate of species $k$, which the species conservation equation (\ref{eqn:gas_specie}) requires, is \cite{chemfoamWiki}
\begin{equation}\label{eqn:omega_k}
    \dot{\omega}_{k} = \sum_{i}^{}\dot{\omega}_{i}(\nu_{k,i}'-\nu_{k,i}'')\quad.
\end{equation}

Hence, for methane combustion, the reaction rates for each species are 
\begin{equation}
\begin{bmatrix}\dot{\omega}_{\ce{CH4}} \\ \dot{\omega}_{\ce{O2}}  \\ \dot{\omega}_{\ce{CO2}}  \\ \dot{\omega}_{\ce{H2O}} \end{bmatrix} = \begin{bmatrix}\dot{\omega}_1(1-0) \\ \dot{\omega}_1(2-0) \\ \dot{\omega}_1(0-1) \\ \dot{\omega}_1(0-2)\end{bmatrix} =\begin{bmatrix}\dot{\omega}_1 \\ 2\dot{\omega}_1 \\ -\dot{\omega}_1 \\ -2\dot{\omega}_1\end{bmatrix} \quad.
\end{equation}

As an inter-specie, \ce{N2} does not participate in chemical reactions. Consequently, 
\begin{equation} \label{eqn:N2}
    Y_{\ce{N2}} =  1 - \sum_{{k}\neq \ce{N2}} Y_{k}
\end{equation}
calculates the mass fraction for \ce{N2} explicitly.

\subsubsection{Turbulence-Chemistry Interaction Model}

Combustion involves complex interactions between fluid dynamics and chemistry, with turbulence playing a key role by enhancing reactant mixing and accelerating reactions. To model the chemical reaction rates for each species in Equation (\ref{eqn:gas_specie}), the Partially Stirred Reactor (PaSR) model calculates the reaction rate using the turbulent, $\tau_\text{t}$, and chemical, $\tau_\text{c}$, time scales, along with the instantaneous reaction rate as follows \cite{chomiak},
\begin{equation} \label{eqn:PaSR}
\overline{\dot{\omega}}_\text{k} = \frac{\tau_\text{c}}{\tau_\text{c} + \tau_\text{t}}\dot{\omega}_\text{k}.
\end{equation}
The turbulent time scale is calculated as a fraction of the turbulence integral time scale ($ \tau_\text{t,I}$),
\begin{equation}\label{eqn:tau_t}
    \tau_\text{t} =C_\text{m} \tau_\text{t,I}  = C_\text{m} \sqrt{\frac{\nu_\text{eff}}{\varepsilon}} ,
\end{equation} 
where  $C_\text{m}$ is the mixing coefficient. \( C_\text{m} \) equals unity for laminar ($Re_\text{t}=0$) and zero for highly turbulent flows ($Re_\text{t}\rightarrow \infty$).  Typically, \( C_\text{m} \) is between 0.01 and 0.3 for turbulence flows; for example, $ C_\text{m} \thickapprox 0.1$ for  $Re_\text{t} \thickapprox 1000$. 

The chemical time scale is the ratio of the total concentration to the production rate of the product species expressed as,
\begin{equation}\label{eqn:tau_c}
    \tau_\text{c} = \sum_\text{i=1}^{I}\frac{c_\text{total}}{\sum_\text{k=1}^\text{K$_\text{products}$}\nu_\text{k,i}\cdot\dot{\omega}_\text{k,i}},
\end{equation} 
 where $c_\text{total}$ denotes the overall concentration of all species and K$_\text{products}$ all product species.

\subsubsection{Radiation Modeling}

\begin{figure}[t]
    \centering
    \includegraphics[width=0.5\textwidth]{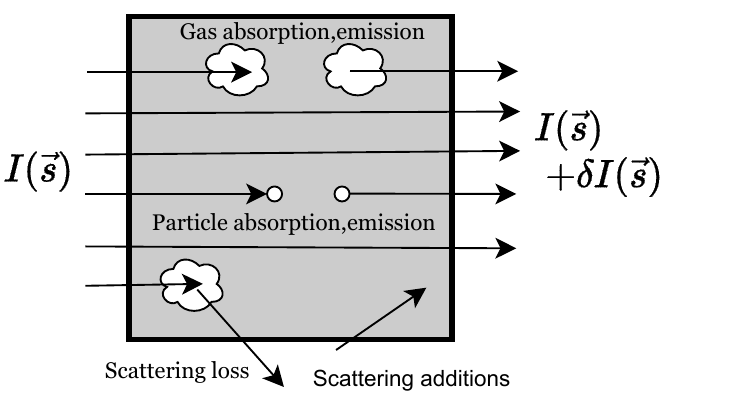}
    \caption{Overview of radiative transfer equation modeling inside a control volume.}
    \label{fig:RTE}
\end{figure}

Due to the high temperature, radiation cannot be neglected and is a main heat source for particles from the fluid field before the flame-front arrives \cite{Liberman2021}. We solve the radiative transfer equation that describes the change in radiation intensity $I(\vec{s})$ in the direction of $\vec{s}$. The conservation of $I$ includes emission and absorption according to the Stefan–Boltzmann law, scattering losses, and absorption by the gas (\ce{CO2} and \ce{H2O}) and particles; see Figure \ref{fig:RTE} for an overview.

This study applies the discrete ordinates radiation model \cite{RTE} because it accounts for gas participation in the radiative transfer equation. The discrete ordinates radiation model discretizes the angular space into a finite set of solid angles, each representing a direction, $\vec{s}$, and solves the radiative transfer equation for these angles.

\subsection{Solid Phase Governing Equations}

\begin{figure*}[tb]
    \centering
    \includegraphics[width=0.9\textwidth]{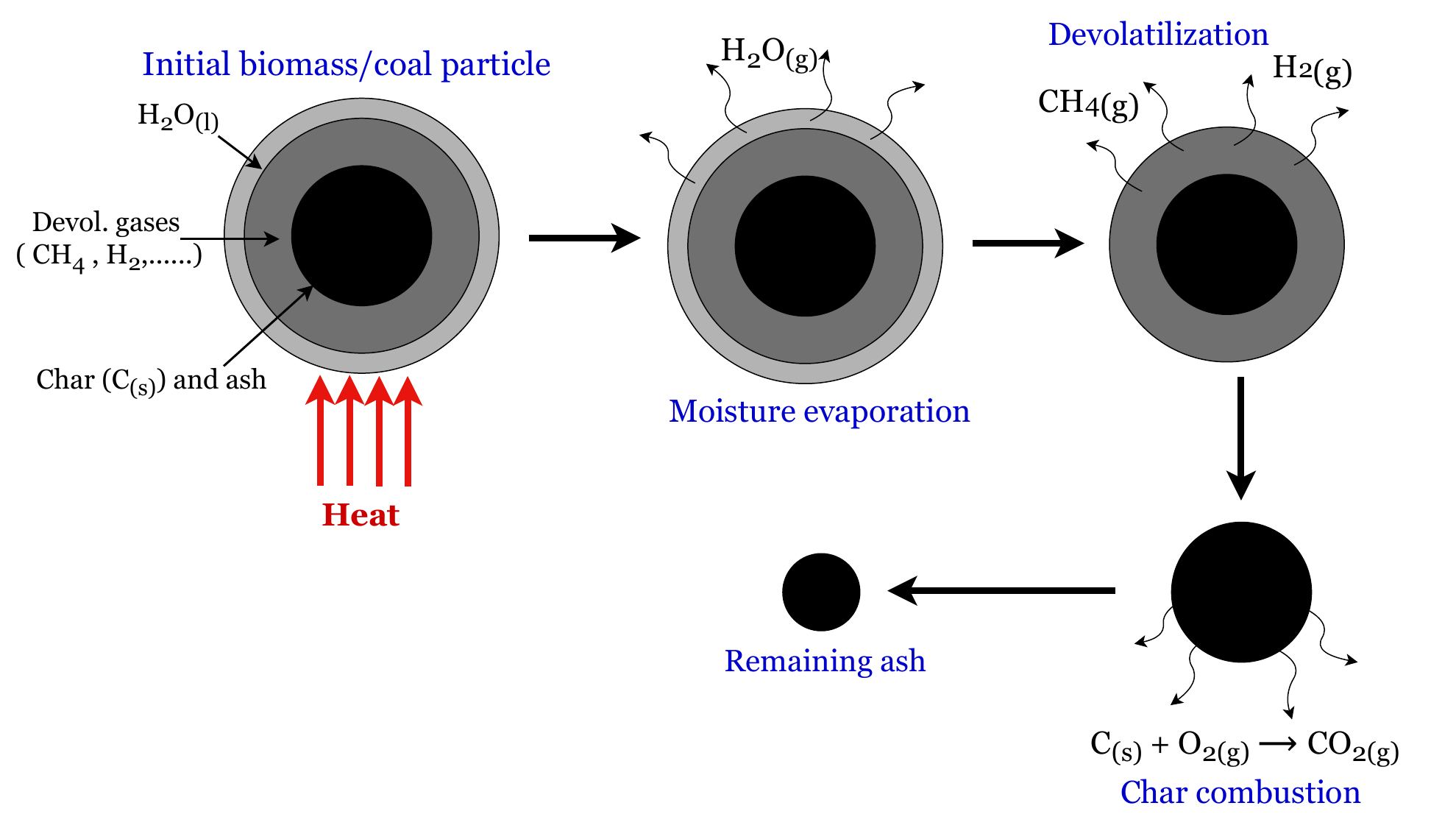}
    \caption{Overview of the biomass/coal particle combustion process.}
    \label{fig:particleprocess}
\end{figure*}

The governing equations for the dispersed solid phase are based on the Lagrangian particle tracking approach. Similar to the gas phase, the solid phase equations also consist of the conservation laws for mass, momentum, and energy,
\begin{align} \label{eqn:particleConti}
    \frac{dm_\text{p}}{dt} &= \frac{dm_\text{evp}}{dt} + \sum\frac{dm_\text{devol,k}}{dt} + \frac{dm_\text{char}}{dt}, \\[6pt]
\label{eqn:particleMomen}
m_\text{p} \frac{d\textbf{u}_\text{p}}{dt} &=  \textbf{F}_\text{d} + \textbf{F}_\text{g}, \\[6pt]
\label{eqn:particleEnergy}
m_\text{p}C_\text{p}\frac{dT_\text{p}}{dt} &=\dot{Q}_\text{conv} + \dot{Q}_\text{radi} + \frac{1}{dt}\bigl\{ \Delta H_\text{evp} + \Delta H_\text{devol} + \Delta H_\text{char}\bigr\}.
\end{align}

According to mass conservation (Equation (\ref{eqn:particleConti})), the combustion of a particle occurs in three main steps. As illustrated in Figure \ref{fig:particleprocess}, an initial particle (such as biomass or coal) consists of char, moisture, and volatile gases. Upon particle heating, moisture evaporates first, followed by the release of volatile gases and, finally, char combustion. These products are the source terms for each species' continuity equation. In the momentum equation (Equation (\ref{eqn:particleMomen})), drag and buoyancy are the external forces on the particle. The energy balance (Equation (\ref{eqn:particleEnergy})) for the particles involves convection, radiation, and latent heat.

\subsubsection{Moisture Evaporation Model}

We assume water to be the only liquid in the particles. 
The Water vapor mass transfer rate by evaporation reads,
\begin{equation}
    \frac{dm_\text{evp}}{dt} = A_\text{p} Ni W_\text{H$_2$O} ,
\end{equation}
where, $A_\text{p}$ is the particle's surface area and $Ni$ is the molar flux of water vapor,
\begin{equation}\label{eqn:vapourFlux}
    \textcolor{black}{Ni = k_\text{c} ( C_\text{s} - C_{\infty})
    = \frac{Sh}{d_\text{p}}D_{\text{H$_2$O}} \biggl(\frac{p_\text{sat}}{R_\text{u}T_\text{s}} - X_{\ce{H2O}} \frac{p}{R_\text{u}T_\text{s}}\biggr) .}
\end{equation}

\textcolor{black}{The water vapor concentration in the bulk gas, \(  C_\text{s} \), is determined based on the saturation pressure, \( p_{\text{sat}} \), of water vapor at the fluid film temperature, \(T_\text{s} \). This temperature is estimated using the rule of thirds for particle and fluid temperatures at the particle surface. The bulk gas water vapor concentration, \( C_{\infty} \), corresponds to a vapor molar fraction of \( X_{\ce{H2O}} \). The mass transfer coefficient, \( k_\text{c} \), is derived from the particle diameter, \( d_\text{p} \), the vapor diffusivity, \( D_{\text{H$_2$O}} \), and the Sherwood number, \( Sh \).}

The Sherwood Number (\(Sh\)), analogous to the Nusselt number in heat transfer, represents the ratio between the total (convective and diffusive) to diffusive mass transfer. According to dimensional analysis, $Sh$ is a function of the Reynolds and the Schmidt ($Sc = \nu/D_\text{H$_2$O})$ numbers. We estimate this function by the Ranz-Marshall correlation \cite{ranz1952evaporation},
\begin{equation}
\label{eqn:Sh}
    Sh = 2 + 0.6\, Re_\text{p}^{1/2} Sc^{1/3}.
\end{equation} 
The particle Reynolds number is given by,
\begin{equation}\label{eqn:Re_p}
Re_\text{p} = \frac{\rho_\text{f} d_\text{p} |\textbf{u}_\text{f} - \textbf{u}_\text{p}|}{\mu_\text{f}}.
\end{equation}

The required heat for moisture evaporation ($\Delta H_\text{evp}$) is calculated from the enthalpy difference between the gas phase and the water vapor at the given pressure and temperature.

\subsubsection {Devolatilization Model}

Devolatilization decomposes volatile matter (gas) from a particle until the remaining char combusts. Thermogravimetric analysis and derivative thermogravimetry determine experimentally $T_\text{devol}$, the temperature at which the devolatilization starts, and $\Delta H_\text{devol}$, the required latent energy \cite{coalWeiguo}.
In this simulation, a first-order kinetic rate model calculates the devolatilized mass of species,
\begin{equation}\label{eqn:devol}
\dfrac{dm_\text{devol,k}}{dt} = k(T) \,(m_\text{devol,k} - m_\text{devol,k,0}) ,
\end{equation}
where \(dm_\text{devol,k}\) is the devolatilized mass , \(m_\text{devol,k}\) is the current mass, and \(m_\text{devol,k,0}\) is the initial mass of species \(k\). The kinetic rate \(k(T)\) is temperature-dependent and is determined by an Arrhenius equation.
%
%

\subsubsection{Surface Reaction Model (Char Combustion)}

After the complete devolatilization, the particle remains as char, which we assume to consist of pure carbon. Then, this char undergoes oxidation,
\begin{equation}\label{eqn:charcon}
\text{C}_\text{(s)} + Sb\text{O$_2$}_\text{(g)} \rightarrow \text{CO$_2$}_\text{(g)}.    
\end{equation}
Therein, the stoichiometric coefficient $Sb$ is set to unity. 

Mass consumption of char is, according to the kinetic-diffusion limited rate model \cite{baum1971},
\begin{equation}\label{eqn:charbombus}
\dfrac{dm_\text{char}}{dt} = - A_\text{p}p_\text{O$_2$}\frac{\mathcal{D}\mathcal{K}}{\mathcal{D}+\mathcal{K}} ,
\end{equation}
where the partial pressure of \ce{O2} is
\begin{equation}
p_\text{O$_2$} = \rho R_u T_\text{f} \biggl(\frac{Y_\text{O$_2$}}{W_\text{O$_2$}}  \biggl),
\end{equation}
the diffusion rate coefficient, 
\begin{equation}
\mathcal{D}= C_1\frac{(T_\text{f} + T_\text{p})^{0.75}}{2d_\text{p}},
\end{equation}
and the kinetic rate coefficient,
\begin{equation}
\mathcal{K} = C_2 \text{exp}({-E_\text{a}}/{R_uT_\text{p}}).
\end{equation}
In these equations, $C_1$, $C_2$, and $E_\text{a}$ are the mass diffusion-limited rate constant, the kinetics-limited rate pre-exponential constant, and the kinetics-limited rate activation energy, respectively.

Due to the reaction, it is necessary to determine the consumption of \ce{O2} and the production of \ce{CO2} and update the gas-phase equations. Hence, the consumption of \ce{O2} equals
\begin{equation}
    \textcolor{black}{dm_\text{O$_2$} = Sb  W_\text{O$_2$}  \frac{dm_\text{char}}{W_\text{c}},}
\end{equation}
and the production of \ce{CO2} is
\begin{equation}
     \textcolor{black}{dm_\text{CO$_2$} = \frac{dm_\text{char}}{W_\text{c}}(W_\text{c} + SbW_\text{O$_2$}).}
\end{equation}

The energy released from char combustion is given by,
\begin{equation}
    \Delta H_\text{char} = h_{\text{ret}}\Delta h_\text{f,CO$_2$}dm_\text{CO$_2$}, 
\end{equation}
where $\Delta h_\text{f,CO$_2$}$ is the formation enthalpy of \ce{CO2}. However, a portion of this total released energy is transferred to the particle itself, apart from the gas phase. This portion is defined by the retention coefficient $h_{\text{ret}}$, which is assumed to be 0.3 for char conversion to \ce{CO2} \cite{boyd1986three}.

\subsubsection{Forces Acting on Solid Particles}

The drag force is given by the expression
\begin{equation}
    \textbf{F}_\text{d} = C_\text{d}A \frac{\rho_\text{f}}{2}|\textbf{u}_f - \textbf{u}_p|(\textbf{u}_f - \textbf{u}_p) ,
\end{equation}
where $A$ is the cross-section area of the particle, and $C_\text{d}$ is the drag coefficient which is calculated by the empirical correlation \cite{Putnam1961,Amsden1989},
\begin{equation}
C_\text{d} = 
\begin{cases} 
\frac{24}{Re_\text{p}}\left(1 + \frac{1}{6}Re_\text{p}^{2/3}\right) & Re_\text{p} \leq 1000 \\ 
0.424 & Re_\text{p} > 1000 .
\end{cases}
\end{equation}

The combined buoyancy and gravitational force reads
\begin{equation}
    \textbf{F}_\text{g} = m_\text{p} \biggl(1 - \frac{\rho_\text{f}}{\rho_\text{p}}\biggl)\textbf{g}.
\end{equation}

\subsubsection{Heat Transfer Model}

This section describes the models for the source terms in the energy conservation equation of the particles (\ref{eqn:particleEnergy}). The convective heat transfer between the fluid and a particle is 
\begin{equation}\label{eqn:q_conv}
    \dot{Q}_\text{conv} = hA_\text{p}(T_\text{f} - T_\text{p}).
\end{equation}
The convective heat transfer coefficient, $h$, is according to the Nusselt number definition, 
\begin{equation}\label{eqn:NU} 
h = \frac{k_\text{f}}{d_\text{p}}Nu ,
\end{equation}
where $k_\text{f}$ is the thermal conductivity of the fluid.
The Nusselt number, $Nu$, is a function of the Reynolds and Prandtl, $Pr$, numbers.
Analogous to Equation~(\ref{eqn:Sh}), the Nusselt number follows the Ranz-Marshall correlation as \cite{ranz1952evaporation}
\begin{equation} \label{eqn:Ranz}
   Nu = 2 + 0.6 Re^{1/2}_\text{p} Pr^{1/3}.
\end{equation}
from the empirical and the particle Reynolds number.

Due to moisture evaporation from the surface, the convective heat transfer is corrected by the `Bird Correction' \cite{bird_transport_2002},
\begin{equation}
     \dot{Q}_\text{conv,corrected} = h\frac{\beta}{e^{\beta} - 1}A_\text{p}(T_\text{f} - T_\text{p}),
\end{equation}
where, $\beta$ is $NiC_{p,\text{H$_2$O}}W_\text{H$_2$O}$  (cf.~Equation~(\ref{eqn:vapourFlux})).

Particles participate in both the absorption and emission of radiation energy. Therefore, the resultant radiative heat transfer,
\begin {equation}
    \dot{Q}_\text{radi} = \dot{Q}_\text{radi,absorption} - \dot{Q}_\text{radi,emission} = A_\text{p}\epsilon_\text{p}\Biggl[\frac{G}{4} - \sigma T_\text{p}^4 \Biggr].
\end{equation}

The radiation model calculates the incident radiation, $G$, for each fluid cell by solving the radiative transfer equation, where $\epsilon_\text{p}$ is the particle emissivity and $\sigma$ denotes the Stefan–Boltzmann constant. 

The energy equation includes three parts of heat transfer from the combustion processes. The enthalpy difference from the moisture evaporation model determines $\Delta H_\text{evap}$. Thermogravimetry experiments determine the value of $\Delta H_\text{devol}$. The latent heat from char combustion, $\Delta H_\text{char}$, is calculated from the formation enthalpy of \ce{CO2} and the mass of \ce{CO2} generated from char combustion. Moisture evaporation and devolatilization absorb energy from the fluid (endothermic process), and char combustion releases heat into the fluid (exothermic process).

\subsection{Coupling between Gas and Particle Phases}

This section describes the link between source terms in the gas-phase governing equations~\eqref{eqn:gas_mass}~to~\eqref{eqn:gas_specie} and the particle-phase governing equations~\eqref{eqn:particleConti}~to~\eqref{eqn:particleEnergy}. 

The rate of a species transferred from the particles $P$ within a volume $dV$ to the gas equals the source term in the gas-phase species transport equation,
\begin{equation}
S_\text{Y,k} = \frac{d}{dV} \sum_i^P \dfrac{dm_\text{i,k}}{dt} ,
\end{equation}
and accordingly, the source term in the gas-phase continuity equation becomes
\begin{equation}
   S_\rho = \sum_k S_\text{Y,k} .
\end{equation}

The sums of the drag and buoyancy forces of the particles inside a volume equal the source terms for the gas phase momentum equation.
Analogously, the sum of the particles' net heat transfer of the three combustion processes is the source term of the gas phase energy equation. 

\section{Set-up, Comparison to Experiments, and Mesh Convergence}

\subsection{\textcolor{black}{Experimental Setup}}

\textcolor{black}{We validate the numerical solver by comparing it to the experimental results of lycopodium dust explosions in the 20~L sphere.  We used 15 g of lycopodium dust, resulting in a nominal dust concentration of 750 g$\text{m}^{-3}$ inside the sphere, and two 1 kJ chemical igniters \cite{SPITZER2025105560}. For the experiments, we apply the setup documented by \cite{JAV}, who explored the distribution of lycopodium dust inside the sphere for various nozzle types. As mentioned in the Introduction, normally, the dust is stored in a small canister and released into the 20~L sphere via a valve. However, we used the "Janovsky" nozzle in the experiments and placed the dust directly in the sphere, which resulted in a more uniform distribution within the sphere.
Further, this approach ensures the total dust mass within the sphere, contrary to rebound or annular nozzles, where some dust can remain trapped within the nozzle and connecting pipes. In other words, we chose an experimental setup and procedure optimized for model validation.}  

 \subsection{Chemical and Material Properties of Lycopodium }

This section documents the chemical and material properties of lycopodium dust and the numerical input for the simulation.
Explosion parameters vary with the precise source of the dust, even if the material is nominally the same. The composition and size distributions can differ for each dust. Additionally, some property trade-offs reduce simulation complexity. 

Explosive dust contains solid material, moisture, and volatile gases. Volatile gases may include \ce{CH4}, \ce{H2}, \ce{CO}, and others. In this study, we assume that the volatile gas is \ce{CH4} and the solid of the particle is pure carbon~(\ce{C})~\cite{RASAM2019747}. A Sartorius MA35 test determined the moisture content twice, yielding values of 3.4 weight-\% and 3.8 weight-\%, which suggests the presence of volatile matter \cite{JAV}.

One reason why we decided to study lycopodium is that its particles have a narrow size distribution. A Malvern Mastersizer3000 measured a mean particle size of $d_{50}$~=~30.6~$\upmu$m, $d_{10}$~=~25.3~$\upmu$m, and $d_{90}$~=~36.9~$\upmu$m. In the simulation, all particles are spherical with a uniform diameter of d50. The material density of lycopodium dust is 1000 kg/m\(^3\), and the specific heat capacity is $C_\text{p}~=~1005$~J/kgK. The devolatilization temperature is $T_\text{dev}$=~483 K, and the devolatilization latent heat is $\Delta H_\text{dev}~=~3.07 \times 10^5$~J/kg \cite{PORTARAPILLO2023112737}.

\subsection{Numerical Set-up}

This section provides key numerical inputs for the simulations.

The geometry is a sphere with a radius of 0.168 m, equivalent to a volume of 20 L. For simplicity, the nozzle, ignition source components inside the sphere, and the dust-filling mechanism are not considered for modeling. Instead, the dust particles are initially placed within the sphere. 

The simulation domain has a single boundary, the wall of the sphere. The wall temperature is maintained at a fixed value of 294 K, corresponding to the water jacket in the experiment. The velocity boundary condition is no-slip, and the pressure is specified as a zero gradient at the wall. Standard wall functions are used for the turbulence parameters ($k$, $\varepsilon$, and $\nu_t$). As initial values, the temperature, pressure, and velocity field are set uniformly to 294~K, 1~bar, and zero, respectively.

Ignition is modeled by adding a temperature patch inside the domain. The ignition parameters are finalized using an iterative approach adopted from Pan~\cite{leobenphd}. Accordingly, the ignition volume consists of two spheres, each with a radius of 4 cm, positioned at the center of the sphere with a center-to-center distance of 8~cm. The temperature of the ignition source is time-dependent, following a first-order polynomial ($T = 2 \times 10^5$\,K/s $t$ + 294 K), with an ignition duration of $t=$ 20 ms as an average value of typical ignition time between 10-30 ms.

The CFL number is set to 0.1, and a first-order Euler time integration scheme is used. The spatial discretization schemes are second-order accurate. The $k-\varepsilon$ turbulence model, commonly used in combustion problems, is selected. The chemical solver uses an implicit Euler method to handle the small time steps. Additionally, a relaxation factor of 0.8 is applied to all fields for simulation stability.

\subsection{Mesh Convergence}
\begin{figure}[t]
\subfloat[]{\includegraphics[width=0.5\textwidth]{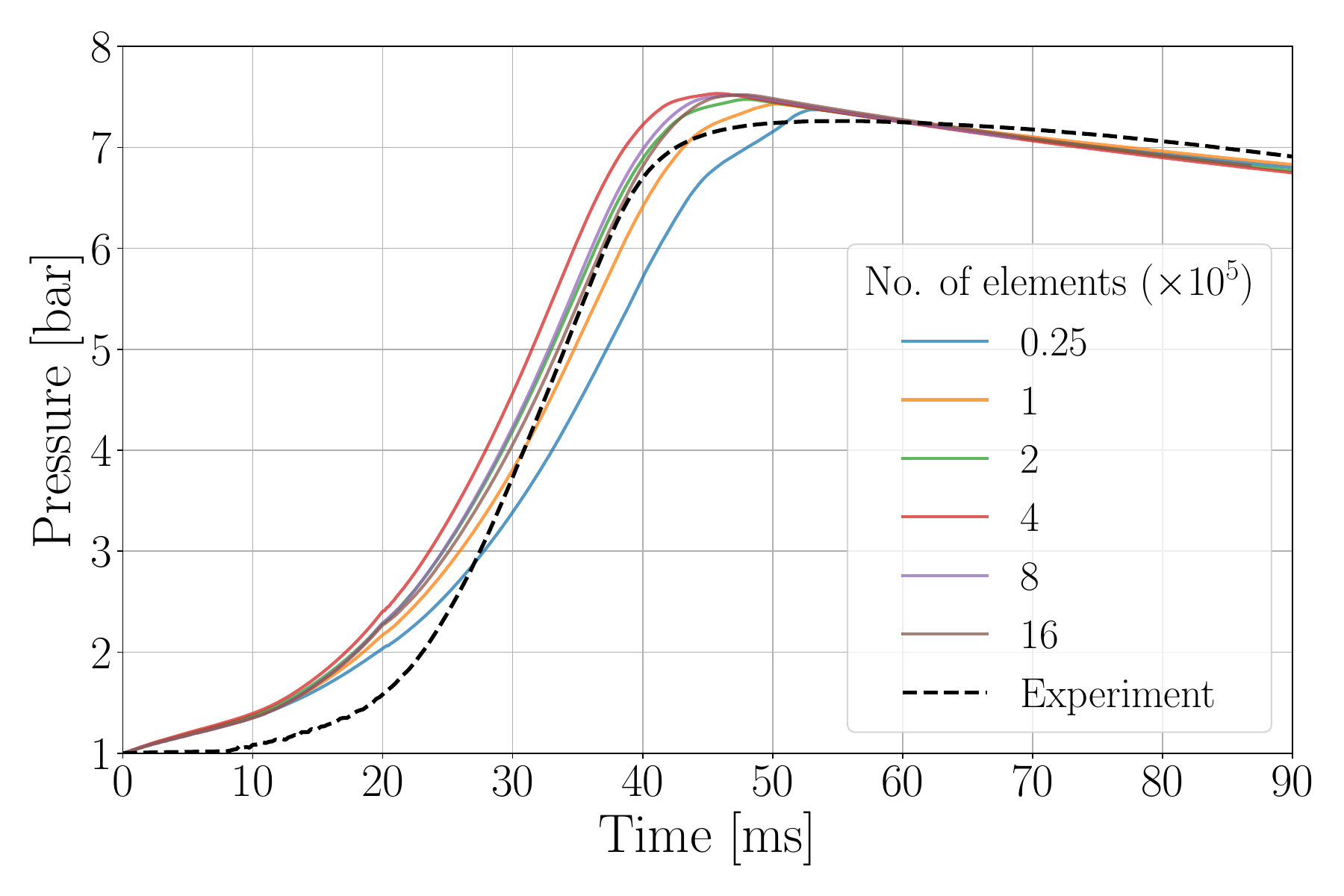}\label{fig:mesh_con_pres}}\\
\subfloat[]{\includegraphics[width=0.5\textwidth]{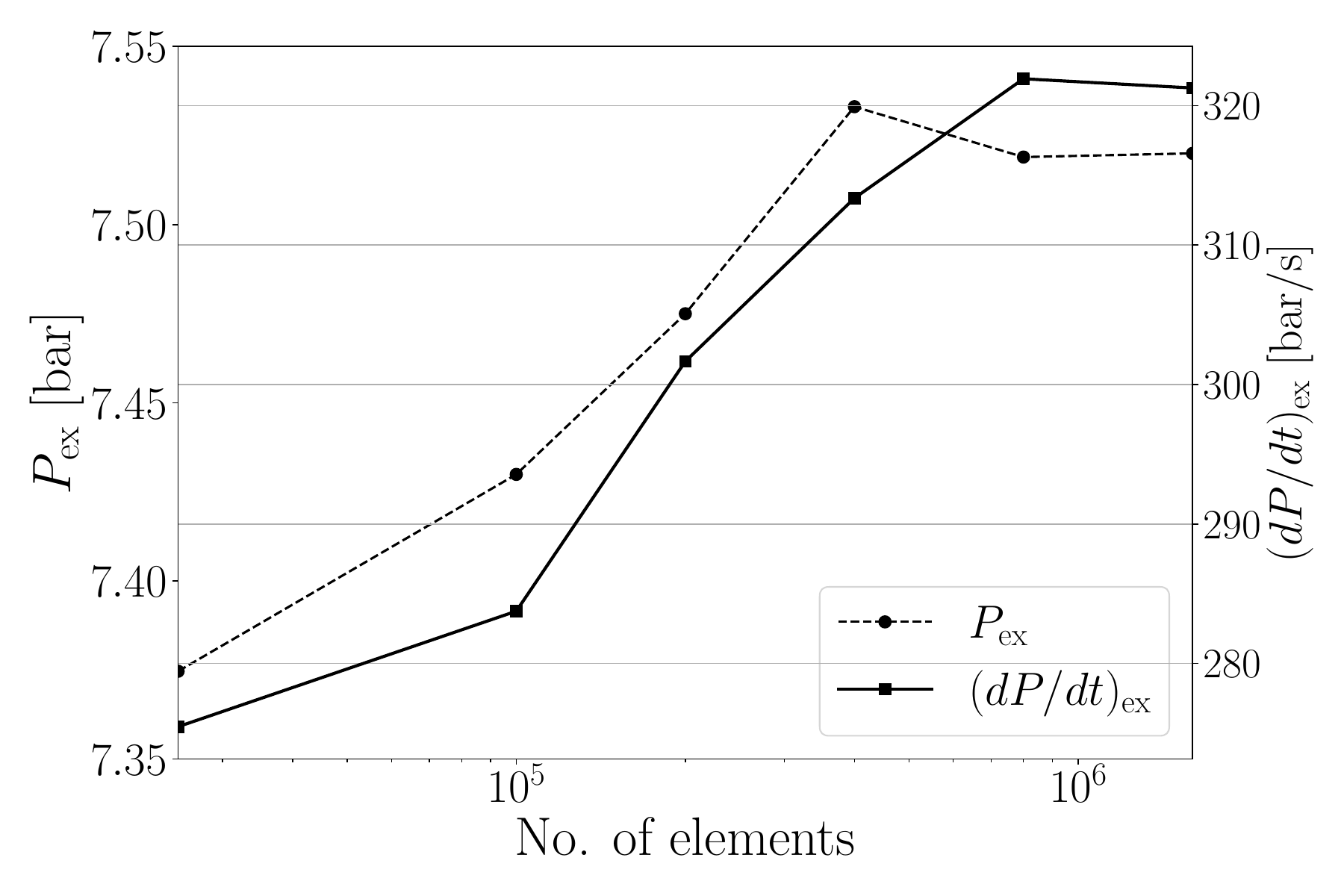}\label{fig:mesh_con_cells}}
\caption{(a) Pressure curves of simulations with different mesh sizes compared to experimental data. (b) $P_\text{ex}$ and $({dP}/{dt})_\text{ex}$ for different mesh sizes.}
\end{figure}

The mesh of the spherical domain consists of 7 blocks, with 1 cubical block at the center and 6 spherical caps surrounding it. All elements are hexagonal, and the mesh is structured. The experiment provides time-resolved pressure data for benchmarking the CFD model. \textcolor{black}{However, it is limited to pressure data only, and experimental results do not include temperature data or combustion products.} For the mesh convergence test, we computed the dust explosion on 6 meshes with element counts ranging from  $0.25\times10^5$ to $16\times10^5$. Figure \ref{fig:mesh_con_pres} compares the pressure developments per mesh to the experimental result. Figure \ref{fig:mesh_con_cells} presents  of $P_\text{ex}$ and $(dP/dt)_\text{ex}$ against the mesh size. Evaluating the computational time and accuracy, we decided to use the mesh with $8\times10^5$ elements, having errors of $3.5\%$ and $2.5\%$ relative to the experimental data for $P_\text{ex}$ and $(dP/dt)_\text{ex}$. Further, we use $5\times10^5$ particles. \textcolor{black}{As the number of elements increases, the numerical results converge.} 

\textcolor{black}{
Modeling errors related to physical phenomena, omitting dust injection and geometrical simplifications, cause a gap between the numerical and experimental results. Particularly, neglecting dust injection and ignition source modeling results in a discrepancy during the ignition phase, while combustion and chemical reaction modeling cause an excessive energy release, leading to a higher $P_\text{max}$ compared to the experimental value. In contrast, the influence of boundary conditions introduces relatively low errors, as evidenced by the small gap between experimental and CFD results during the settling phase after $P_\text{max}$.}

\section{Particle Homogeneity Analysis and Results}

\subsection{Initial Particle Distribution}

Lycopodium dust achieves a comparatively uniform distribution before ignition, which is why it was selected for the benchmark as a model dust. \textcolor{black}{The low terminal settling velocity, caused by the rough surface texture of the particles, along with their nearly mono-sized distribution, ensures consistent behavior in air, leading to uniform particle distribution \cite{ECKHOFF2003199}}. Nevertheless, as discussed in the introduction, particles are generally more concentrated toward the wall region than uniformly distributed. Therefore, this study investigates how the particle distribution affects the safety characteristics.

To this end, we start the simulations using different initial particle distributions, where the coefficient $\Phi$ characterizes each distribution's the radial homogeneity.
First, we generate a uniform distribution of particles in a unit sphere, each particle having a distance to the unit sphere's center of $r_\mathrm{uni}$.
Then, we scale the radial distance of the particles to the sphere's center by $r_\Phi=R r_\mathrm{uni}^\Phi$, where $R=$~0.168~m is the sphere's radius.
Thus, $\Phi$ varies between 0 and 1, where $\Phi=0$ implies that all particles are placed at the wall and $\Phi=1$ that the particles are uniformly distributed.

\begin{figure}[tb]
\subfloat[]{\includegraphics[width=0.48\textwidth]{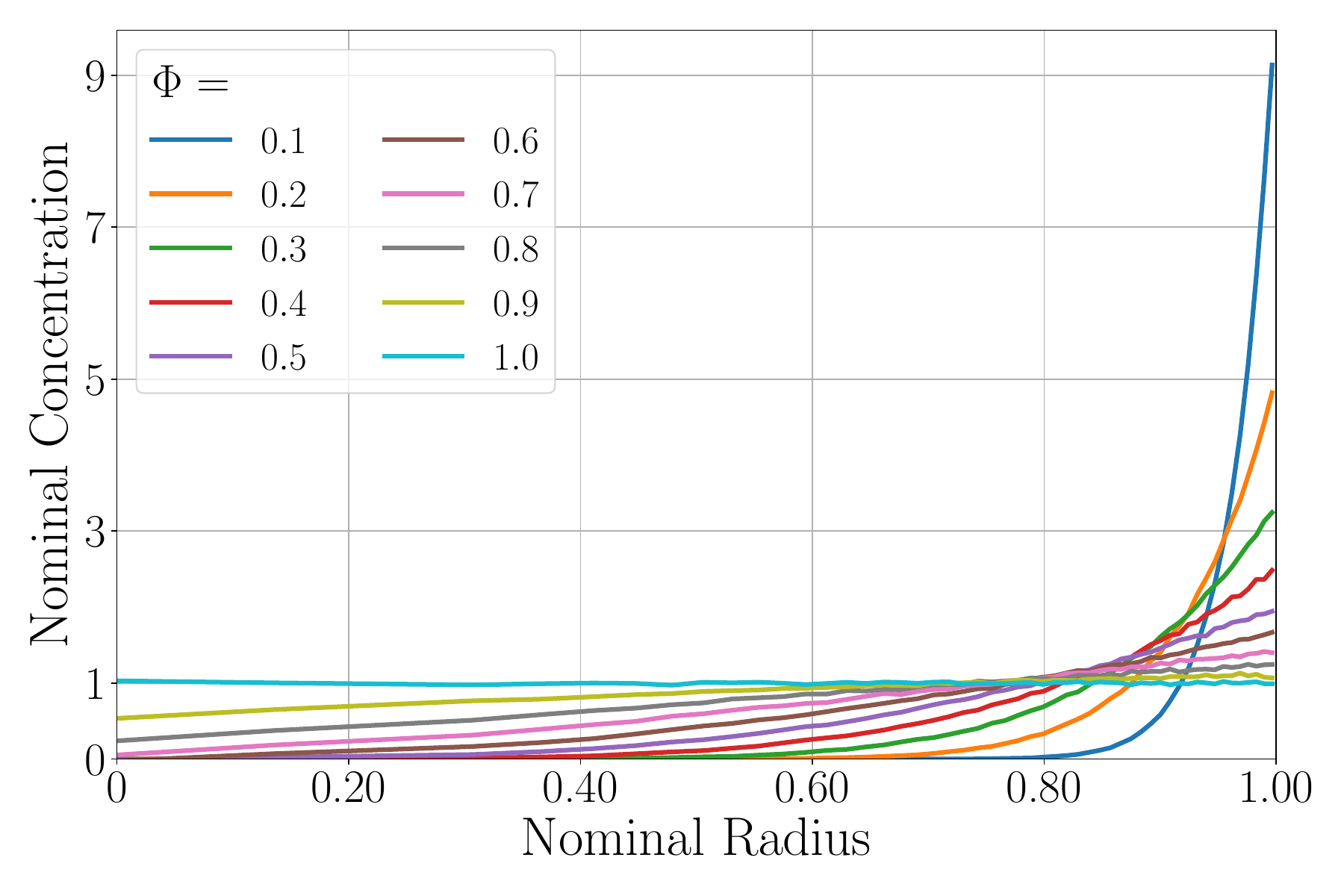}\label{fig:con_initial}}\\
\subfloat[]{\includegraphics[width=0.23\textwidth]{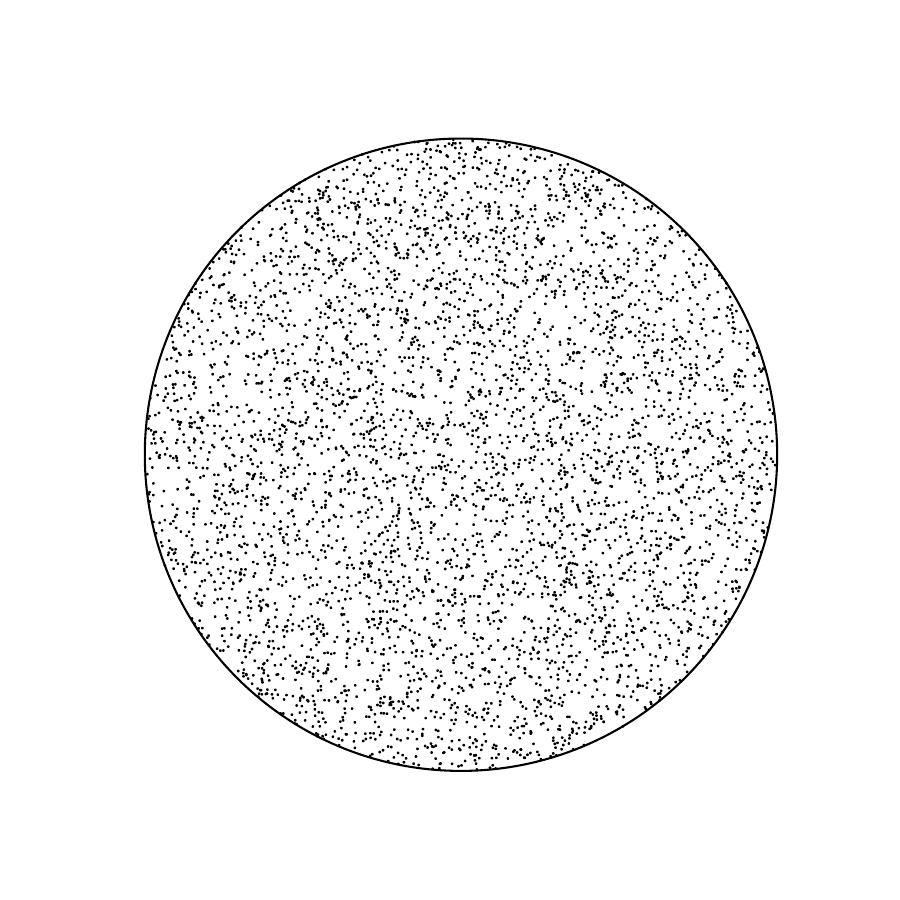}\label{fig:phi_1}}~
\subfloat[]{\includegraphics[width=0.23\textwidth]{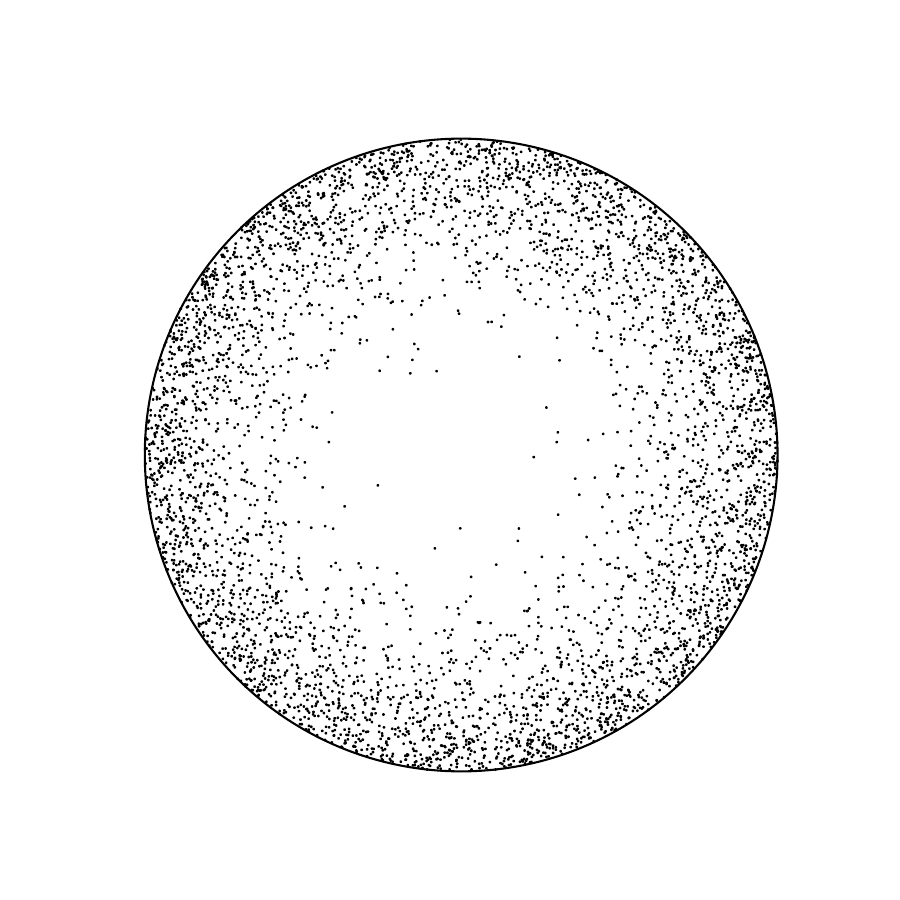}\label{fig:phi_0_5}}
\caption{Initial particle distribution (a) along the radial coordinate, for (b) $\Phi = 1$, and (c) $\Phi = 0.5$.}
\end{figure}

Multiple simulations explored the effect of particle radial distribution on explosions by varying the particle distributions by $\Phi = {0.1, 0.2, 0.3, \ldots, 1}$ while keeping all other parameters constant. Figure~\ref{fig:con_initial} shows the initial nominal concentration (relative to the uniform distribution concentration) of the particles along the radial coordinate normalized by the sphere's radius. For the $\Phi=1$ (Figure~\ref{fig:phi_1}), the concentration is uniform along the radius. For $\Phi=0.5$ (Figure~\ref{fig:phi_0_5}), the particles concentrate at the wall. For the $\Phi = 0.1$, most particles consolidate in the outer $20\%$ of the radius, with a concentration 9 times higher near the wall than the uniform distribution.


\subsection{Safety characteristics}

\begin{figure*}[tb]
    \centering
    \subfloat[]{\includegraphics[width=0.49\textwidth]{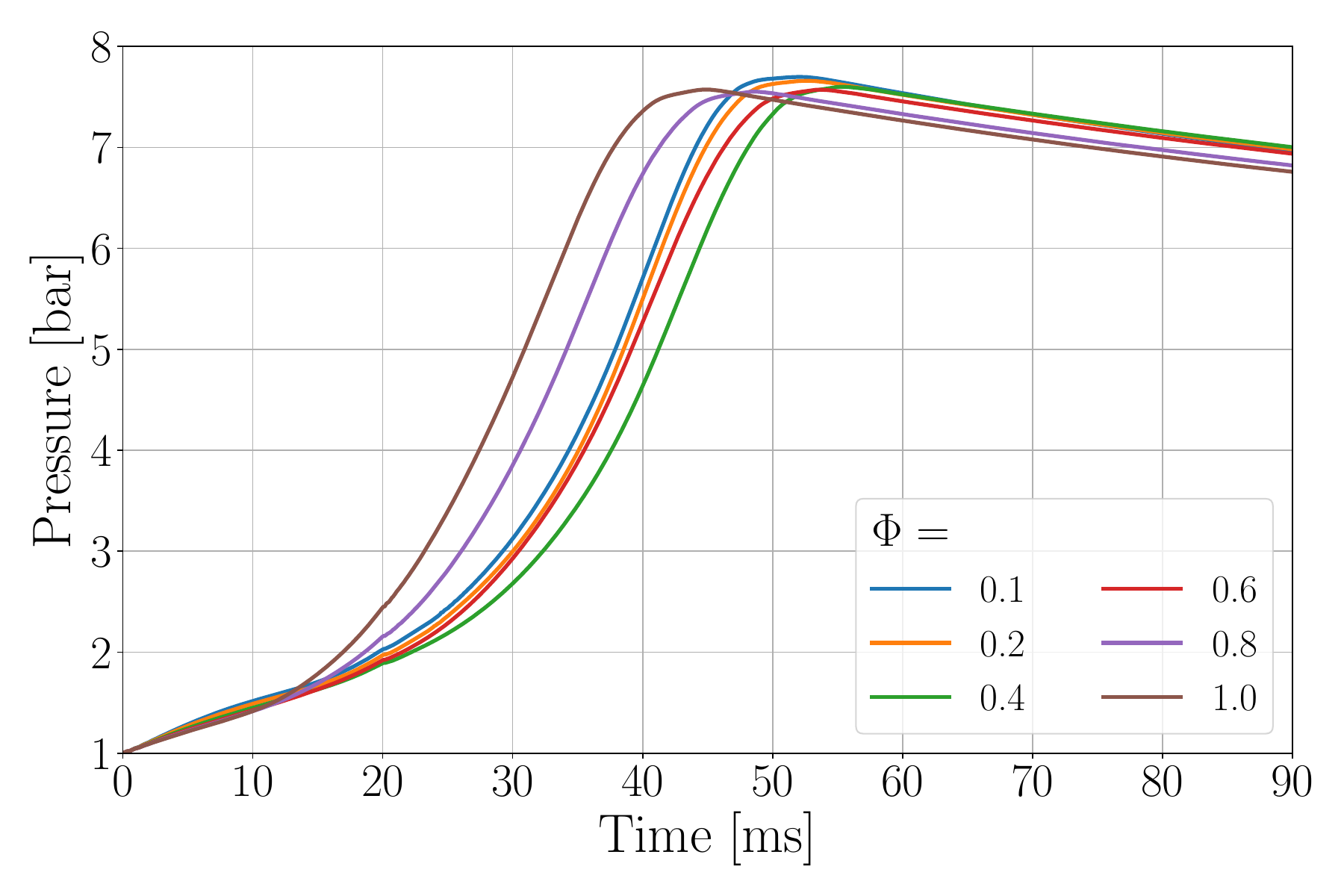}\label{fig:pVst_phi}}~~
    \subfloat[]{\includegraphics[width=0.49\textwidth]{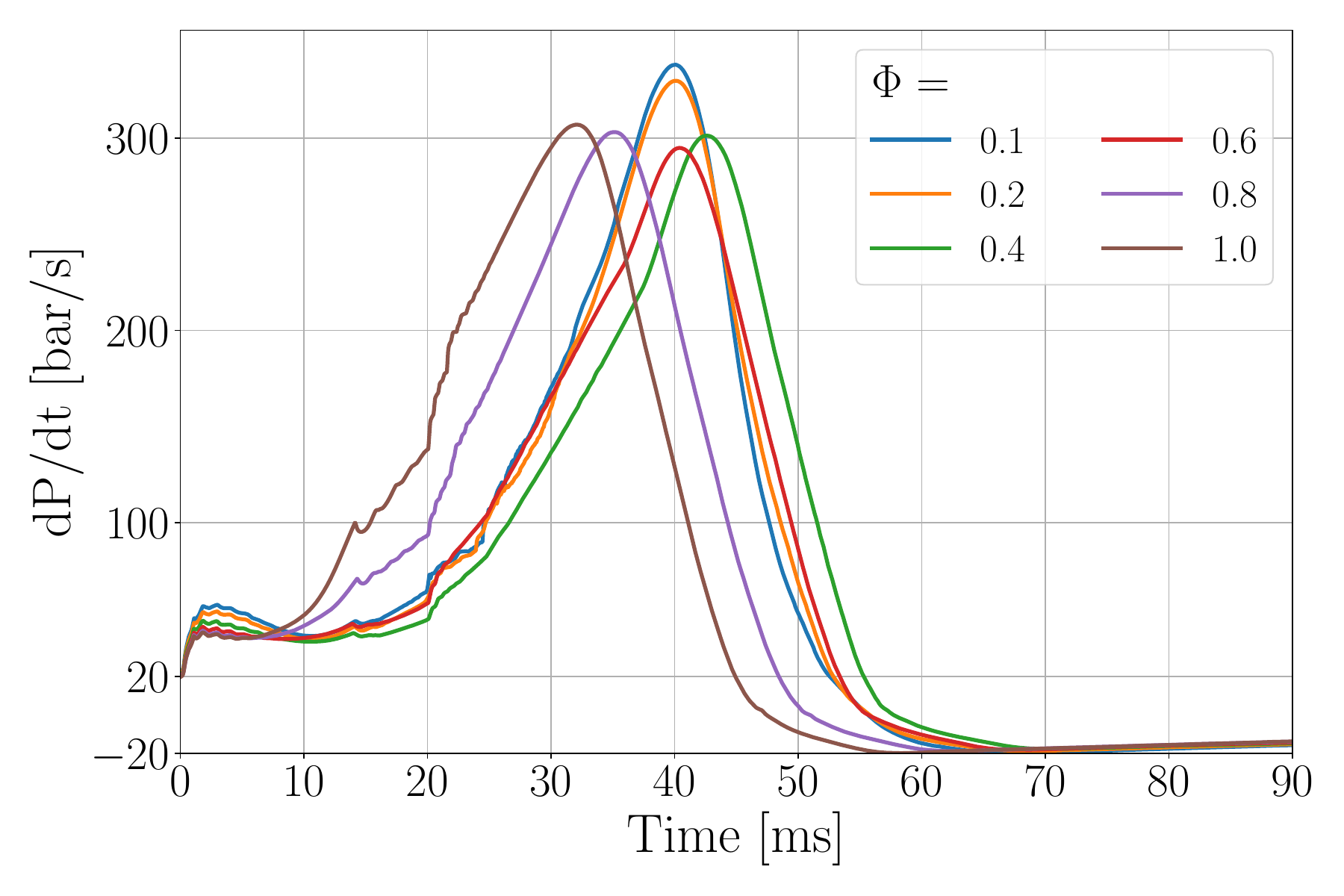}\label{fig:dpdtVst_phi}}
    \caption{(a) Pressure and (b) pressure gradient in the sphere vs. time for different particle distributions.}
    \label{fig:Vst_phi}
\end{figure*}

Figure \ref{fig:Vst_phi} shows the pressure and pressure gradient development curves for each particle distribution. When $\Phi$ decreases from 1 to 0.4, it takes longer to reach the explosion pressure. For a further decrease below $\Phi = 0.4$, the trend reverses, and the explosion pressure is reached faster. When particles are uniformly distributed and ignition starts, the ignition kernel covers some particles, initiating an explosion. Consequently, the flame front propagates rapidly, driven by the chain reactions of volatile gas combustion; the pressure and the rate of pressure rise increase. 

In contrast, a void exists at the center (near the igniter) when particles are closer to the walls. Upon ignition, the flame front takes more time to reach and ignite the particles, thus delaying the explosion. When particles are concentrated near the wall ($1>\Phi>0.4$), the flame front propagates quickly toward the wall without significant combustion occurring in the midsection. This reduces the overall explosion intensity. However, if the particles are excessively concentrated near the wall ($\Phi<0.4$), the flame front encounters a much higher particle concentration in that region, leading to an increased explosion rate, see Figure (\ref{fig:dpdtVst_phi}). 

Figure \ref{fig:dpdtVst_phi} shows the pressure gradient variation with time, indicating the explosion's intensity. Initially, higher $\Phi$ yield a rapid explosion rate, with the time to reach $(dP/dt)_\text{ex}$ being the shortest for the uniform distribution. Afterward, the time to reach $(dP/dt)_\text{ex}$ increases with $\Phi$. However, due to the higher explosion rate near the wall, the time decreases again for $\Phi=0.1$ and $0.2$. The lowest $(dP/dt)_\text{ex}$ occurs when $\Phi= 0.7$, while the highest occurs when $\Phi = 0.1$. The non-linear relationship between $\Phi=0.1$ and $(dP/dt)_\text{ex}$ results from the competition between the reduced number of particles encountered by the flame front for low $\Phi$ while propagating through the sphere, which reduces $(dP/dt)_\text{ex}$, and encountering a higher particle concentration close to the wall, which increases $(dP/dt)_\text{ex}$.

\begin{figure}[tb]
    \centering
    \includegraphics[width=0.48\textwidth]{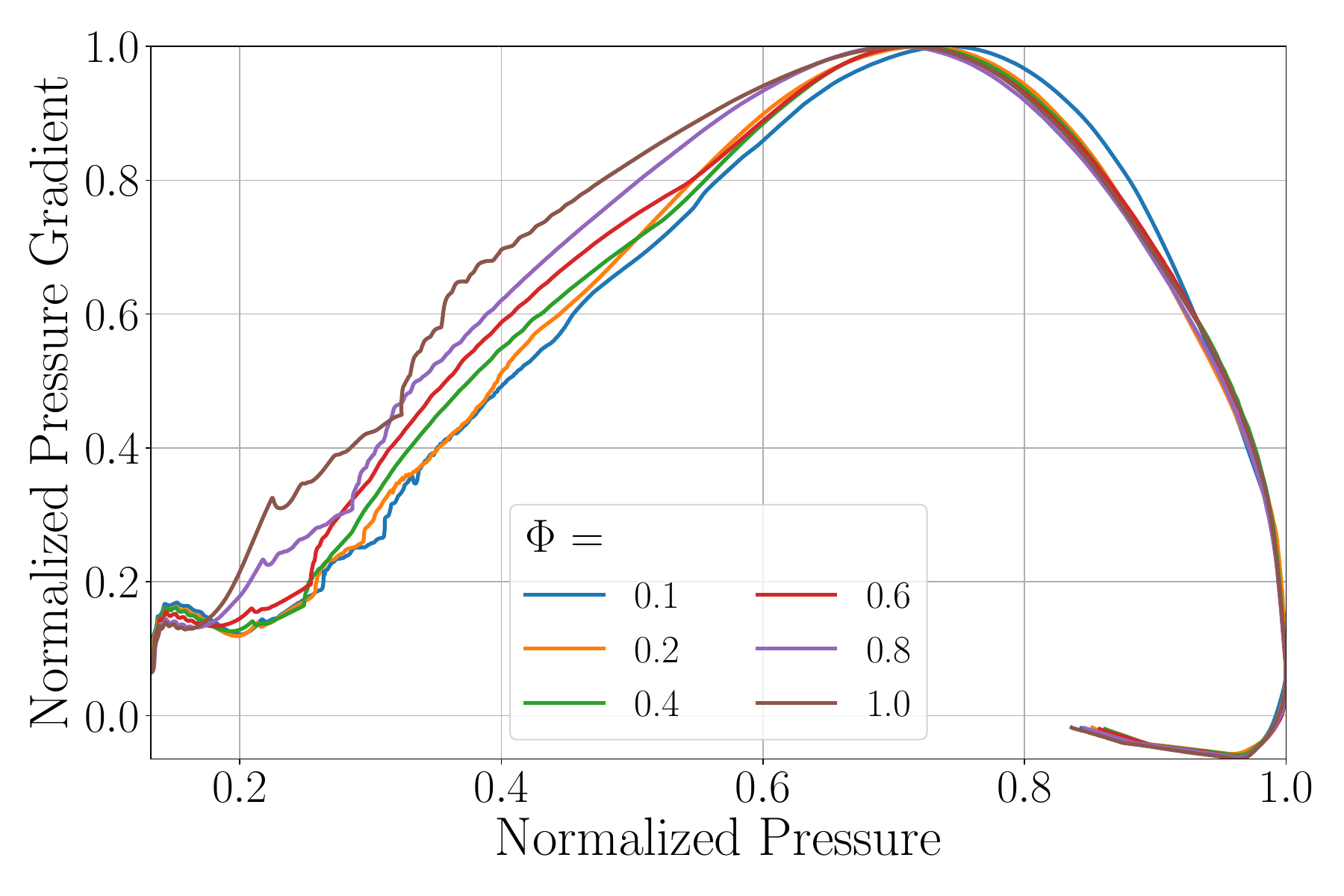}
    \caption{Normalized pressure gradient $\Bigl(\frac{dP/dt}{(dP/dt)_\text{ex}}\Bigr)$ vs normalized pressure ($P/P_\text{ex}$).}
    \label{fig:NdpdtVsNp}
\end{figure}

Figure \ref{fig:NdpdtVsNp} visualizes the pressure and pressure gradient, normalized by their maximum value for each simulation.
As suggested by \cite{Bauwens2024}, this provides an understanding of the explosion parameters independent of time. Nearly all distributions reach their $(dP/dt)_\text{ex}$ when the pressure approaches approximately $72\%$ of its $P_\text{ex}$. Upon reaching $(dP/dt)_\text{ex}$, all distributions behave similarly: $P_\text{ex}$, $(dP/dt)_\text{ex}$ becomes negative, suggesting that the explosion completes, with the rate of pressure decrease being nearly equal for all distributions.

\begin{figure}[tb]
    \centering
    \includegraphics[width=0.48\textwidth]{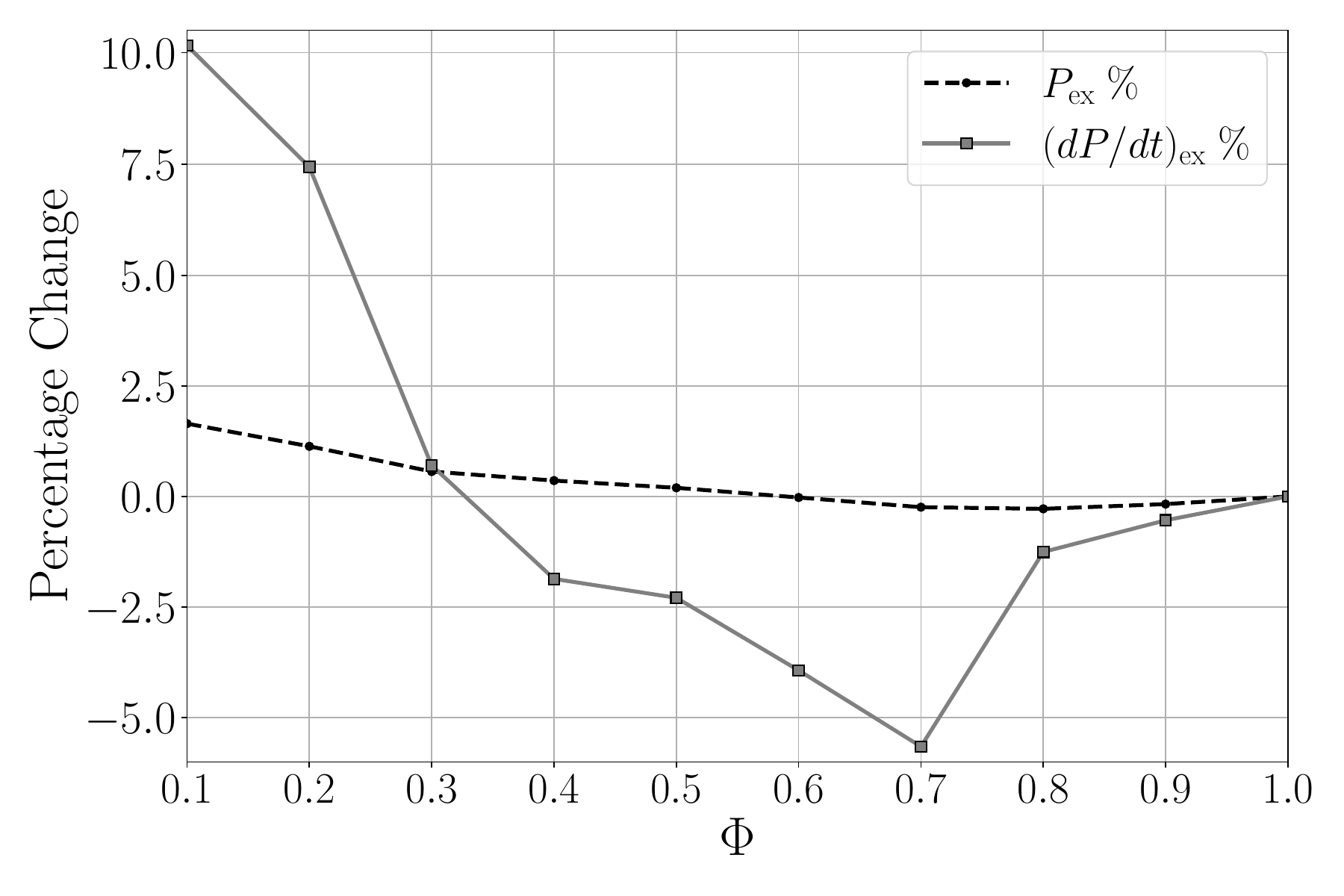}
    \caption{Percentage of variation of $P_\text{ex}$ and $(dP/dt)_\text{ex}$ compared to uniform distribution for each $\Phi$ value.}
    \label{fig:maxpmaxdpdt}
\end{figure}

Figure \ref{fig:maxpmaxdpdt} compares the explosion parameters to the uniform distribution, which means $P_\text{ex}$ and $(dP/dt)_\text{ex}$ as percentages relative to $\Phi=1$. As $\Phi$ decreases, both $P_\text{ex}$ and $(dP/dt)_\text{ex}$ decrease and then increase beyond a certain point. Finally, for $\Phi = 0.1$, $P_\text{ex}$ and $(dP/dt)_\text{ex}$ exceed the uniform distribution by $1.75\%$ and $10\%$, respectively. The minimum $P_\text{ex}$ occurs for $\Phi=0.8$, and $(dP/dt)_\text{ex}$ reaches its minimum for $\Phi=0.7$. This observation implies that moderate particle concentration near the wall ($\Phi = 0.7$) causes a less severe explosion than a uniform distribution.

\textcolor{black}{
To understand why $\Phi=0.1$ causes the highest explosion pressure, we investigate in the next two sub-sections the heat dissipation through the wall and the combustion products.}

\textbf{}

\subsection{Heat Dissipation Through the Wall}

In the experiment, a water jacket is maintained around the sphere at a constant temperature of 294 K to prevent the wall from overheating. Similarly, the wall temperature is set to a constant value in simulations, as described in the boundary condition section. Consequently, the energy released from explosions inside the sphere dissipates through the wall. According to the first law of thermodynamics, the total energy released by the dust combustion and ignition equals the gas' internal energy rise, which is proportional to the pressure rise, and the heat loss through the wall. Thus, a thermally well-insulated sphere, respectively adiabatic boundary conditions in the simulation, would result in a higher $P_\text{ex}$.

\begin{figure}[tb]
\subfloat[]{\includegraphics[width=.48\textwidth]{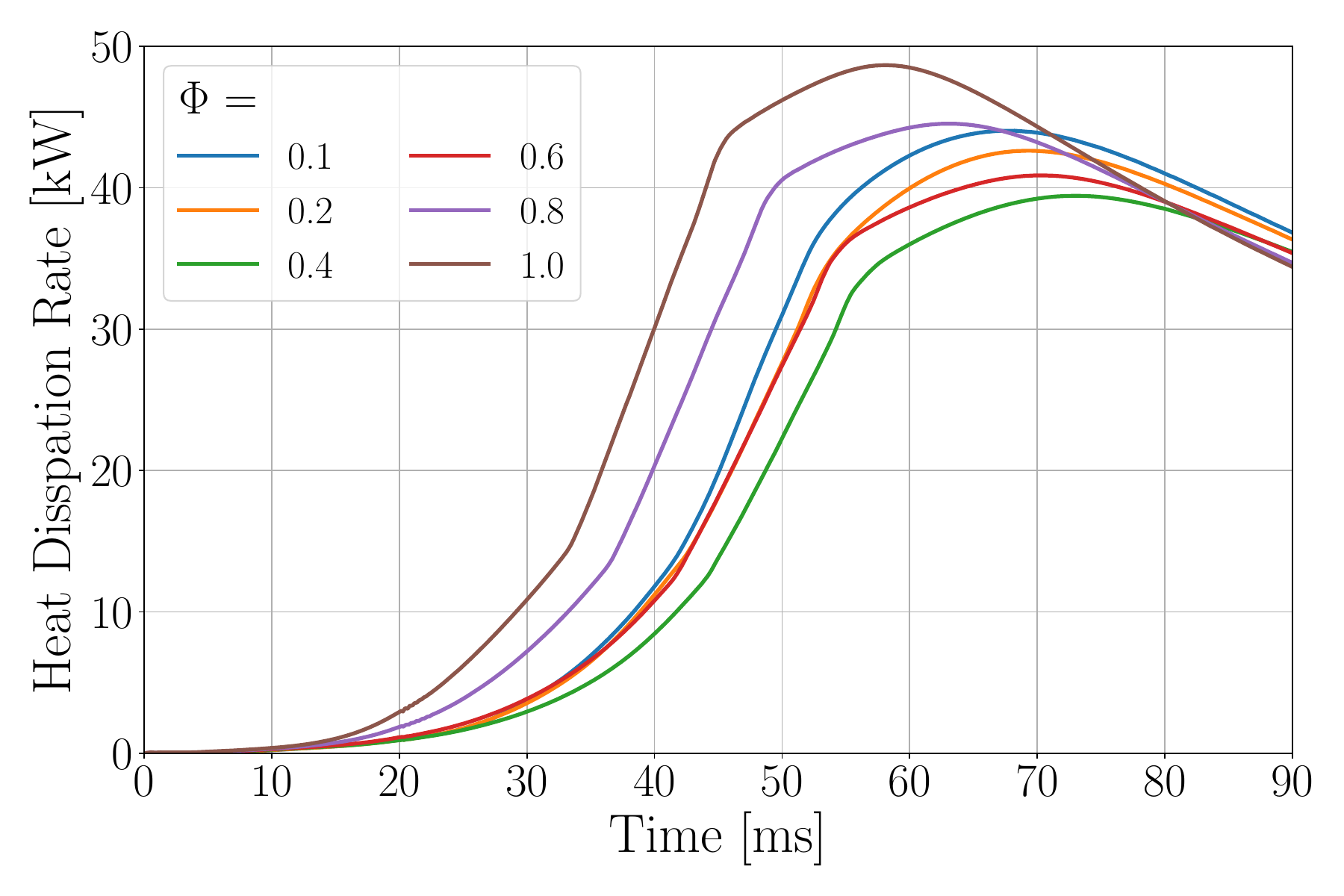}\label{fig:heatlossplot}}\\
\subfloat[]{\includegraphics[width=.48\textwidth]{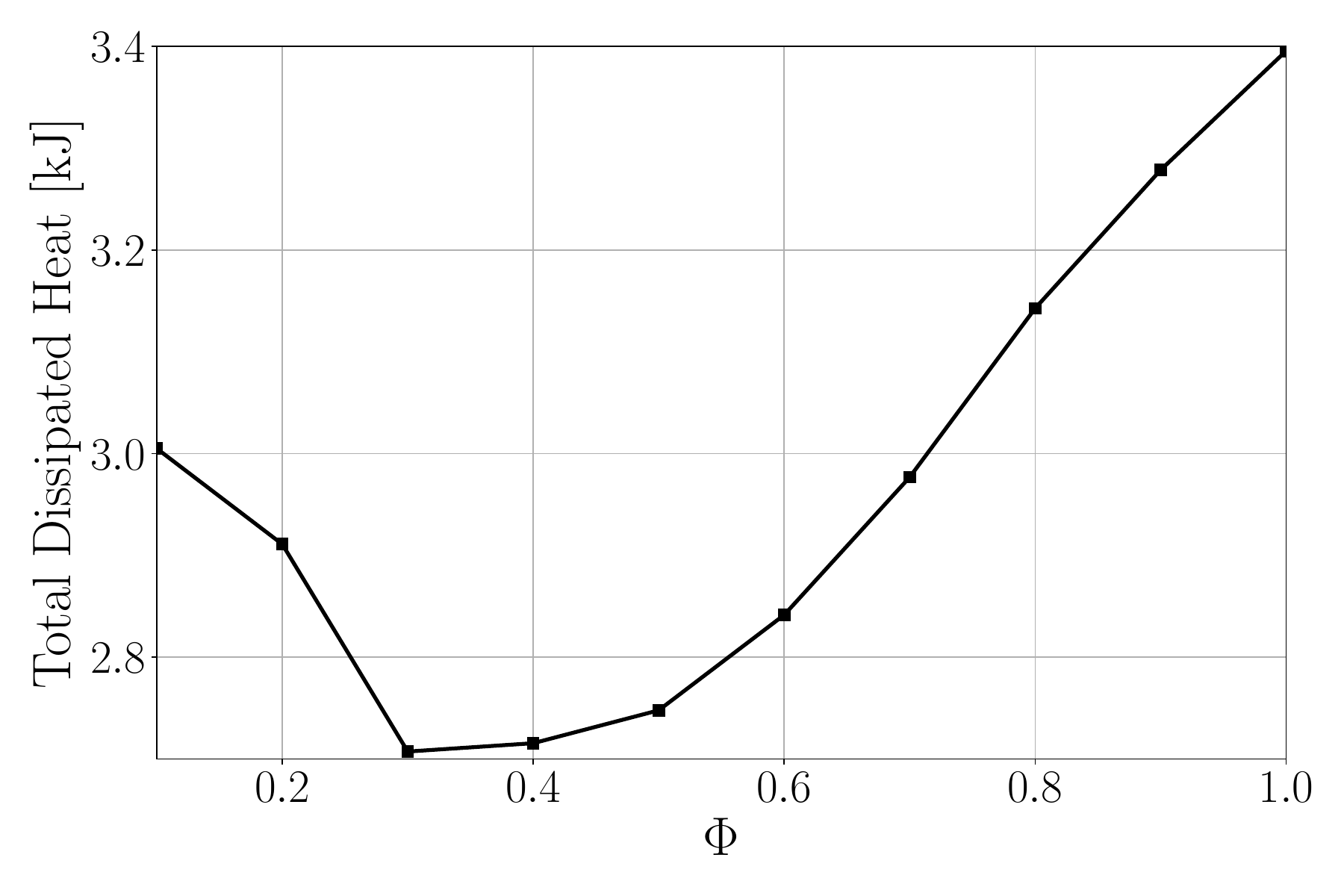}\label{fig:totalheatloss}}
\caption {(a) Heat dissipation rate through the wall over time and (b) total dissipated heat through the wall}
\end{figure}

We calculated the heat loss through the wall by integrating the total heat flux, conductive and radiative, over the wall's surface area.
Figure \ref{fig:heatlossplot} shows the heat dissipation rate through the wall for each distribution. Since the temperature gradient drives the heat loss at the wall, it has a trend similar to the pressure development in Figure \ref{fig:pVst_phi}. Figure \ref{fig:totalheatloss} shows the total heat dissipation during the complete explosion for each distribution. When particles distribute uniformly, they start combusting as soon as ignition starts. Therefore, the internal temperature and the heat flux through the wall increase. However, high particle concentration on the wall delays the combustion. Unburnt particles in the preheat zone absorb the combustion heat, reducing heat loss through the wall. Nevertheless, further increasing the near-wall concentration enhances energy dissipation through the wall. By balancing all of these factors, the lowest energy dissipation rate and total heat loss through the wall occurs for $\Phi=0.3$.

\subsection{Combustion Products}

\begin{figure}[tb]
\includegraphics[width=.48\textwidth]{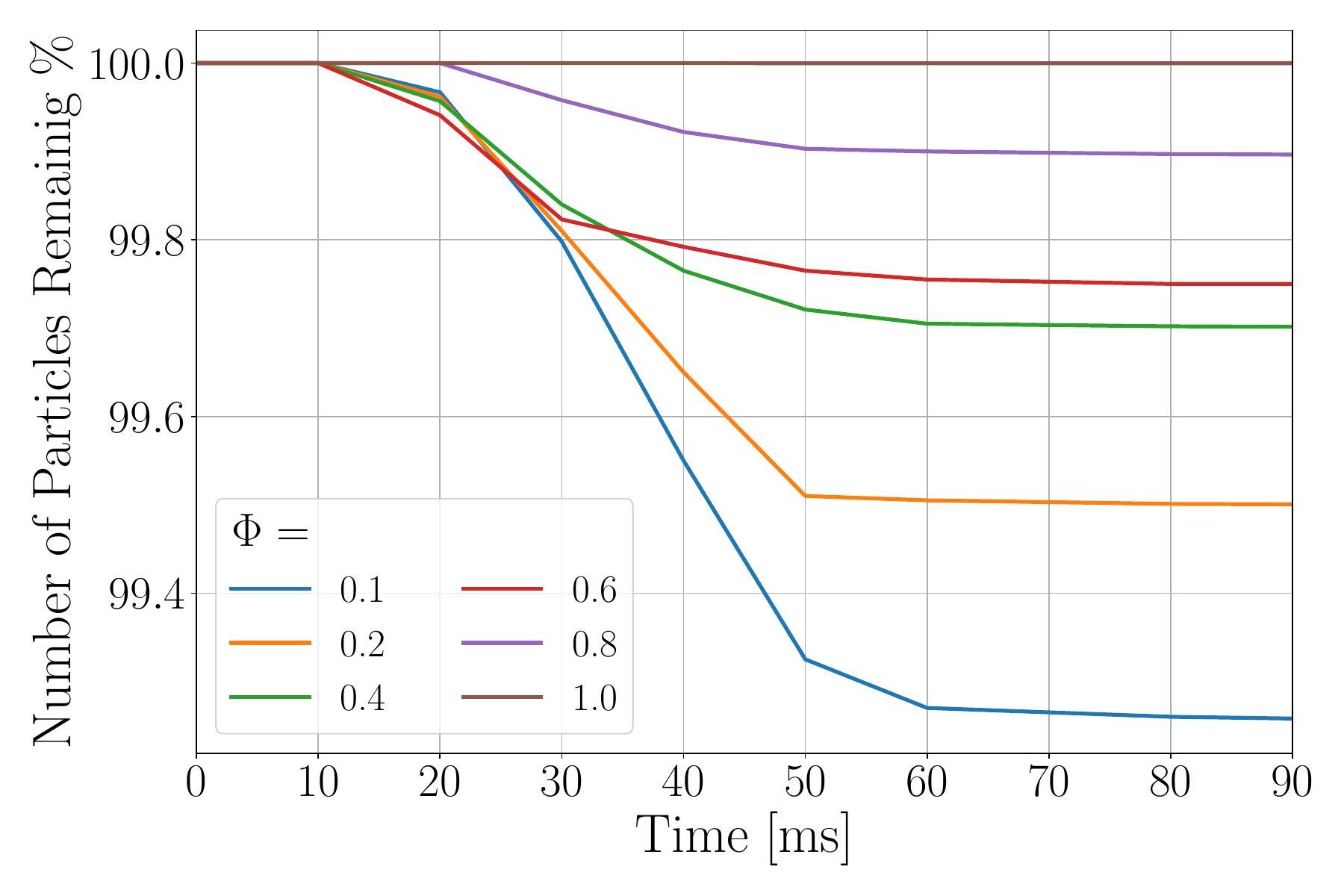}
\caption{Percentage of remaining particles.}
\label{fig:partic_left}
\end{figure}

After moisture evaporation and devolatilization, the solid carbon mass undergoes oxidation. Figure \ref{fig:partic_left} illustrates the percentage of particles remaining inside the sphere over time as a percentage of the initial number of particles. For a uniform distribution, all particles remain while $\Phi$ decreases, and the rate at which particles vanish increases. This could be due to the explosion of particles experiencing a high combustion rate near the sphere's surface.

\begin{figure}
\includegraphics[width=.48\textwidth]{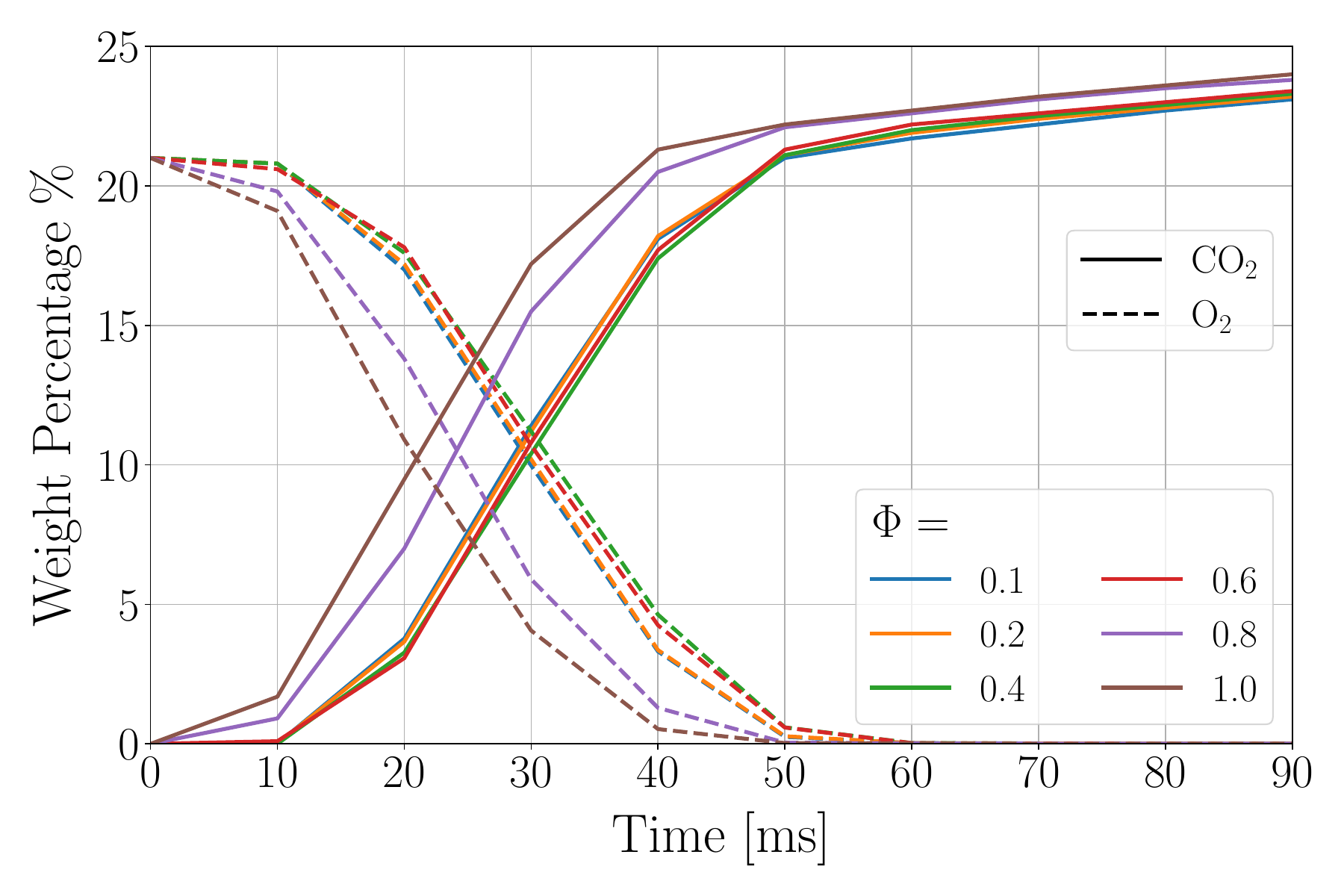}
\caption{Weight of \ce{CO2} and \ce{O2} percentage variation}
\label{fig:CO2O2_con}
\end{figure}

Initially, the air-filled sphere contains only \ce{O2} and \ce{N2}, with molar fractions of 0.21 and 0.78, respectively. Since \ce{N2} is an inert species, it does not participate in reactions. Ultimate end products after all multi-step reactions are \ce{CO2} and \ce{H2O}. Production of \ce{CO2} and consumption of \ce{O2} indicate the amount of fuel or particles burned. Figure \ref{fig:CO2O2_con} shows some distributions' concentration variations of \ce{CO2} and \ce{O2}. After 60 ms, the entire \ce{O2} is consumed for all distributions. However, \ce{CO2} production continues in intermediate reaction steps. For the uniform distribution, the rate at which \ce{O2} is consumed, and the production rate of \ce{CO2} are lower compared to the distribution where particles are near the wall. The findings highlight the significant influence of particle distribution on combustion dynamics, particularly demonstrating how particle positioning within the sphere affects the rates of \ce{O2} consumption and \ce{CO2} production.

\section{Conclusion}

This study investigated the influence of particle distribution on the safety characteristics using a 20~L sphere. The results indicate that explosions are more intense when particles accumulate near the chamber walls than a uniform particle distribution despite a delay in the initial explosion. For the most non-uniform distribution studied ($\Phi=0.1$), the maximum explosion pressure ($P_\text{ex}$) and the maximum rate of pressure rise ($(dP/dt)_\text{ex}$) reached their highest values, with increases of 1.75$\%$ and 10.1$\%$, respectively, compared to the uniform distribution. Conversely, the lowest values for $P_\text{ex}$ and $(dP/dt)_\text{ex}$ were observed at $\Phi=0.7$, showing reductions of 0.25$\%$ and 5.6$\%$, respectively, relative to the uniform distribution. These findings suggest that locally high particle concentrations enhance explosion intensity, whereas moderate concentrations result in lower intensities than a uniform distribution. The observations of heat loss rates and species concentrations further support this behavior. 

\textcolor{black}{The $P_\text{max}$ and $(dP/dt)_\text{max}$ values measured from the 20 L sphere during a particular explosion test were based on the assumption of a uniform particle distribution. However, in reality, the particle distribution is not uniform, leading to variations in safety characteristics. Therefore, further study is required to quantify the necessary corrections to safety characteristics based on actual particle distribution.} This study could be extended to other volumes like the standard cubic meter sphere or other geometries, such as cylindrical chambers that model silos, and dust conveying pipes, to explore further strategies for minimizing explosion damage in process industries.

\section*{CRediT authorship contribution statement}
K. Weerasekara: Writing – review \& editing, Writing – original draft, Visualization, Software, Methodology, Conceptualization. S. H. Spitzer: Writing – review \& editing, Software. S. Zakel: Writing – review \& editing, Supervision. H. Grosshans: Writing – review \& editing, Supervision, Software, Funding acquisition.


\section*{Acknowledgments}
This project has received funding from the European Research Council (ERC) under the European Union’s Horizon 2020 research and innovation program (Grant Agreement No. 947606 PowFEct).

\section*{Data availability}
Data will be made available on request.

\appendix

\section{Nomenclature}
The following table lists the common variables that are not defined in the text.
\renewcommand{\arraystretch}{1.3} 

\bigskip
\begin{tabular}{cll}
    \multicolumn{3}{c}{{\textbf{Variables}}} \\ 
    $A$ & Area & [m$^2$] \\
    $\alpha$ & Thermal diffusion coefficient & [m$^2$/s]] \\
    $D$ & Mass diffusion coefficient & [m$^2$/s] \\
    $d$ & Diameter & [m] \\
    $\varepsilon$ & Turbulence dissipation rate & [m$^2$/s$^3$]\\
    \textbf{g} & Gravitational acceleration & [m/s$^2$] \\
    $h$ & Specific enthalpy & [J/kg] \\
    $H$ & Enthalpy & [J] \\ 
    $k$ & Specific kinetic energy, & [J/kg], \\
        & Turbulent kinetic energy  & [m$^2$/s$^2$] \\
    $m$ & Mass & [kg] \\
    $p$ & Pressure & [Pa] \\
    $Pr$ & Prandtl number & [-] \\
    $\dot{Q}$ & Heat transfer rate & [W] \\
    $R_\text{u}$ & Universal gas constant & [J/(molK)] \\
    ${\mathbf{R}}_f$ & Reynolds stress tensor & [N/m$^2$] \\
    $Re$ & Reynolds number & [-] \\
    $\rho$ & Density & [kg/m$^3$] \\
    $T$ & Temperature & [K] \\
    $t$ & Time & [s] \\
    \boldmath{$\tau$} & Shear stress tensor & [N/m$^2$]\\
    \textbf{u} & Velocity & [m/s] \\
    $\dot{\omega}$ & Chemical reaction rate & [mol/(m$^3$s)] \\

    \end{tabular}
    \label{tab:variables}

\begin{tabular}{cll}
    \multicolumn{3}{c}{\textbf{Subscripts}}\\ 
    conv & Convection & \\
    devol & Devolatilization &\\
    eff & Effective (molecular + turbulent) & \\
    evp & Evaporation & \\
    f & Fluid &\\
    k & Species &\\
    p & Particle & \\
    radi & Radiation & \\ 
    react & Reaction & \\
    t & Turbulent & \\
\end{tabular}

    \label{tab:subscripts}

\bibliographystyle{elsarticle-num} 
\bibliography{ref.bib}

\end{document}